\documentclass[aps,prc,twocolumn,superscriptaddress,showpacs,preprintnumbers,amsmath,amssymb]{revtex4}

\usepackage{graphicx}

\begin{document}


\def\GSI{GSI Helmholtzzentrum f\"{u}r Schwerionenforschung GmbH, D-64291 Darmstadt,
Germany}
\def\MOSCOW{Institute for Nuclear Research, Russian Academy of Sciences,
117312 Moscow, Russia}
\def\KONYA{Department of Physics, University of Sel\c{c}uk, 42079 Konya,
Turkey}
\def\FIAS{Frankfurt Institute for Advanced Studies, J.W. Goethe University,
D-60438 Frankfurt am Main, Germany}
\def\KURCH{Kurchatov Institute, Russian Research Center, 123182 Moscow, Russia}
\def\IPNO{Institut de Physique Nucl{\'e}aire, IN2P3-CNRS et Universit{\'e}, F-91406
Orsay, France}
\def\JAGU{M. Smoluchowski Institute of Physics, Jagiellonian University, Pl-30059 Krak\'ow, Poland}
\def\MILANO{Istituto di Scienze Fisiche, Universit\`a degli Studi and INFN, I-20133 Milano, Italy}
\def\LPC{LPC, IN2P3-CNRS, ISMRA et Universit{\'e}, F-14050 Caen, France}
\def\SACLAY{DAPNIA/SPhN, CEA/Saclay, F-91191 Gif-sur-Yvette, France}
\def\GANIL{GANIL, CEA et IN2P3-CNRS, F-14076 Caen, France}
\def\IFJ{H. Niewodnicza{\'n}ski Institute of Nuclear Physics,
Pl-31342 Krak{\'o}w,
Poland}
\def\CATANIA{Dipartimento di Fisica e Astronomia-Universit\`a and INFN-Sezione CT and LNS, 
I-95123 Catania, Italy}
\def\MSU{Department of Physics and Astronomy and NSCL, MSU, East Lansing, MI 48824, USA}
\def\WARSAW{A.~So{\l}tan Institute for Nuclear Studies, Pl-00681 Warsaw, Poland}
\newcommand{\goo}{\,\raisebox{-.5ex}{$\stackrel{>}{\scriptstyle\sim}$}\,}
\newcommand{\loo}{\,\raisebox{-.5ex}{$\stackrel{<}{\scriptstyle\sim}$}\,}

\title{Isospin dependent multifragmentation of relativistic projectiles}

\affiliation{\GSI}
\affiliation{\KONYA}
\affiliation{\MOSCOW}
\affiliation{\FIAS}
\affiliation{\KURCH}
\affiliation{\IPNO} 
\affiliation{\JAGU}
\affiliation{\MILANO}
\affiliation{\SACLAY}
\affiliation{\GANIL}
\affiliation{\IFJ}
\affiliation{\CATANIA}
\affiliation{\MSU}
\affiliation{\WARSAW}

\author{R.~Ogul}        \affiliation{\GSI}\affiliation{\KONYA}
\author{A.S.~Botvina}       \affiliation{\GSI}\affiliation{\MOSCOW}
\author{U.~Atav}        \affiliation{\KONYA}
\author{N.~Buyukcizmeci}    \affiliation{\KONYA}
\author{I.N.~Mishustin}     \affiliation{\FIAS}\affiliation{\KURCH}
\author{P.~Adrich}              \affiliation{\GSI}  
\author{T.~Aumann}              \affiliation{\GSI} 
\author{C.O.~Bacri}             \affiliation{\IPNO} 
\author{T.~Barczyk}             \affiliation{\JAGU} 
\author{R.~Bassini}             \affiliation{\MILANO}  
\author{S.~Bianchin}            \affiliation{\GSI} 
\author{C.~Boiano}              \affiliation{\MILANO}   
\author{A.~Boudard}             \affiliation{\SACLAY}  
\author{J.~Brzychczyk}          \affiliation{\JAGU}   
\author{A.~Chbihi}              \affiliation{\GANIL}
\author{J.~Cibor}               \affiliation{\IFJ}
\author{B.~Czech}               \affiliation{\IFJ} 
\author{M.~De~Napoli}           \affiliation{\CATANIA}
\author{J.-\'{E}.~Ducret}       \affiliation{\SACLAY}
\author{H.~Emling}              \affiliation{\GSI}
\author{J.D.~Frankland}           \affiliation{\GANIL}
\author{M.~Hellstr\"{o}m}       \affiliation{\GSI}
\author{D.~Henzlova}            \affiliation{\GSI}
\author{G.~Imm\`{e}}            \affiliation{\CATANIA} 
\author{I.~Iori}\thanks{Deceased} \affiliation{\MILANO}  
\author{H.~Johansson}           \affiliation{\GSI} 
\author{K.~Kezzar}              \affiliation{\GSI}
\author{A.~Lafriakh}            \affiliation{\SACLAY}
\author{A.~Le~F\`evre}          \affiliation{\GSI}
\author{E.~Le~Gentil}           \affiliation{\SACLAY}
\author{Y.~Leifels}             \affiliation{\GSI}
\author{J.~L\"{u}hning}         \affiliation{\GSI}
\author{J.~{\L}ukasik}          \affiliation{\GSI}\affiliation{\IFJ} 
\author{W.G.~Lynch}             \affiliation{\MSU}
\author{U.~Lynen}               \affiliation{\GSI} 
\author{Z.~Majka}               \affiliation{\JAGU}  
\author{M.~Mocko}               \affiliation{\MSU}
\author{W.F.J.~M\"{u}ller}      \affiliation{\GSI}
\author{A.~Mykulyak}            \affiliation{\WARSAW}          
\author{H.~Orth}                \affiliation{\GSI}
\author{A.N.~Otte}              \affiliation{\GSI}
\author{R.~Palit}               \affiliation{\GSI}
\author{P.~Paw{\l}owski}        \affiliation{\IFJ}
\author{A.~Pullia}              \affiliation{\MILANO}
\author{G.~Raciti}\thanks{Deceased} \affiliation{\CATANIA}
\author{E.~Rapisarda}           \affiliation{\CATANIA} 
\author{H.~Sann}\thanks{Deceased} \affiliation{\GSI}
\author{C.~Schwarz}             \affiliation{\GSI}
\author{C.~Sfienti}             \affiliation{\GSI}
\author{H.~Simon}               \affiliation{\GSI}
\author{K.~S\"{u}mmerer}        \affiliation{\GSI}
\author{W.~Trautmann}           \affiliation{\GSI}
\author{M.B.~Tsang}             \affiliation{\MSU}
\author{G.~Verde}               \affiliation{\MSU}
\author{C.~Volant}              \affiliation{\SACLAY} 
\author{M.~Wallace}             \affiliation{\MSU}
\author{H.~Weick}               \affiliation{\GSI}
\author{J.~Wiechula}            \affiliation{\GSI}
\author{A.~Wieloch}             \affiliation{\JAGU} 
\author{B.~Zwiegli\'{n}ski}     \affiliation{\WARSAW}

\date{\today}

\begin{abstract}
The $N/Z$ dependence of projectile fragmentation at relativistic energies has been 
studied with the ALADIN forward spectrometer at the GSI Schwerionen Synchrotron (SIS). 
Stable and radioactive Sn and La beams with an incident energy of 600 MeV 
per nucleon have been used in order to explore a wide range of isotopic compositions.
For the interpretation of the data, calculations with the statistical multifragmentation 
model for a properly chosen ensemble of excited sources were performed.
The parameters of the ensemble, representing the variety of excited spectator nuclei expected 
in a participant-spectator scenario, are determined 
empirically by searching for an optimum reproduction of the measured fragment charge 
distributions and correlations. An overall very good agreement is obtained.
The possible modification of the liquid-drop parameters of the fragment description in the hot
freeze-out environment is studied, and a significant reduction of the symmetry-term coefficient
is found necessary to reproduce the mean neutron-to-proton ratios $\langle N\rangle /Z$ 
and the isoscaling parameters of $Z \le 10$ fragments. 
The calculations are, furthermore, used to address 
open questions regarding the modification of the surface-term coefficient at freeze-out, 
the $N/Z$ dependence of the nuclear caloric curve, 
and the isotopic evolution of the spectator system between its formation during the 
initial cascade stage of the reaction and its subsequent breakup.
\end{abstract}

\pacs{25.70.Pq,25.70.Mn,21.65.Ef}

\maketitle

\section{INTRODUCTION}
\label{sec:int}

Multifragmentation is a universal phenomenon occurring when a large
amount of energy is deposited in a nucleus. It has been observed in
nearly all types of high-energy nuclear reactions induced by hadrons,
photons, and heavy ions (for reviews, see Refs.~\cite{gross90,SMM,dyntherm,viola06}).
At low excitation energies, the produced nuclear system can be treated as a compound 
nucleus~\cite{Bohr} which decays via evaporation of light particles
or fission. However, at high excitation energy, possibly accompanied by
compression during the initial dynamical stage of the 
reaction~\cite{bertsch_siemens83,debois87,friedman90}, 
the system will expand to subsaturation densities, thereby becoming unstable, 
and will break up into many fragments. 

Multifragmentation has been shown to be
a fast process with characteristic times around 100 fm/$c$ or less
(see, e.g., Refs.~\cite{beau00,Karna,verm09}).
Nevertheless, as evident from numerous analyses of experimental data,
a high degree of equilibration can be reached in these reactions.
In particular, chemical equilibrium among the produced fragments is very likely 
while kinetic equilibration may not be reached~\cite{xi97}. 
Statistical models were found very suitable for describing the measured
fragment yields \cite{gross90,SMM,Botvina90,Botvina95,EOS,MSU,INDRA,FASA,Dag}. 
Even though equilibration may not occur within the individual system during the primary 
reaction stage the processes are, apparently, of such complexity that dynamical constraints 
beyond the conservation laws do not restrict the statistical population of the
available partition space.  

Taking multifragmentation into account is crucially
important for a correct description of fragment production in high-energy 
reactions. Depending on the chosen projectile, it is responsible for 10 to 50\% 
of the reaction cross section.
Multifragmentation, furthermore, opens a
unique possibility for investigating the phase diagram of nuclear
matter at temperatures $T \approx 3-8$~MeV and densities around
$\rho \approx 0.1-0.3 \rho_0$ ($\rho_0 \approx 0.15$ fm$^{-3}$ denotes
the nuclear saturation density). These equilibrium conditions coincide with those of the nuclear
liquid-gas coexistence region. Very similar conditions are realized in stellar matter during 
the expansion phase of supernova explosions \cite{ishi03,Botvina04,Botvina05}.
The study of multifragmentation thus receives a strong astrophysical motivation from its 
relation to stellar evolution and the formation of nuclei, occurring at densities
near or below $\rho_0$ and over a wide range of isotopic asymmetries. The characteristics
of multifragmentation as a function of the neutron-to-proton ratio $N/Z$ of the disintegrating
system are, therefore, of particular interest. 

Previous ALADIN experimental data have provided
extensive information on charged fragment production in multifragmentation
reactions~\cite{kreutz,schuett96}. In particular, they have
demonstrated a "rise and fall" of multifragmentation with excitation 
energy~\cite{ogilvie91,hubele91}, and they have shown that the temperature remains nearly
constant, around $T\sim$~6~MeV, during this process~\cite{xi97,poch95,traut07}, as predicted
within the statistical approach~\cite{bond85}.
The observed large fluctuations of the fragment multiplicity
and of the size of the largest fragment in the transition region from a
compoundlike decay to multifragmentation have been interpreted as manifestations of the
nuclear liquid-gas phase transition in small systems (see, e.g., 
Refs.~\cite{Dag,campi88,gulm06,chaud07} and references given therein).

In this article, the study of multifragmentation is extended to isotopic effects. 
New data for projectile fragmentation at 600 MeV/nucleon are presented
together with their statistical analysis within the framework of the statistical 
multifragmentation model~\cite{SMM}. The data were obtained during the ALADIN experiment 
S254~\cite{sfienti_prag,traut08,sfienti09} 
in which a maximum range in $N/Z$ was explored by using primary stable beams as well as 
secondary radioactive beams, up to the presently achievable limit on the neutron-poor side.
The projectile isotopic compositions are expected to be 
essentially preserved during the initial stages of the collision,
so that the study can fully benefit from the asymmetry of available secondary beams.

The mass of the produced spectator systems varies considerably with the impact parameter
and is closely linked to the transferred excitation energy. 
The theoretical analysis is, therefore, based on calculations for a full ensemble of 
spectator systems intended to reflect these properties of spectator fragmentation.
The parameters describing the ensemble of excited nuclei 
are chosen to obtain a best 
possible reproduction for a set of experimental observables. We concentrate on observables 
describing the partition degrees of freedom and on the information they carry regarding the 
fragment properties at the chemical freeze-out. 
It is shown that a very satisfactory description of the $N/Z$ dependence of multifragmentation
can be obtained within the statistical framework, including that 
of the freeze-out temperatures~\cite{sfienti09}. The reduction of the symmetry term coefficient 
in the low-density freeze-out environment, reported for $^{12}$C-induced reactions on 
$^{112,124}$Sn isotopes~\cite{LeFevre}, is confirmed and found necessary to 
reproduce the measured mean neutron-to-proton ratios $\langle N \rangle /Z$ and the
isoscaling parameters of intermediate-mass fragments.

\section{EXPERIMENTAL DETAILS}
\label{sec:exp}

The ALADIN experiment S254, conducted in 2003 at the 
GSI Schwerionen Synchrotron (SIS), 
was designed to study isotopic effects in projectile fragmentation at 
relativistic energies. Besides stable $^{124}$Sn, neutron-poor secondary Sn 
and La beams were used to extend the range of isotopic compositions 
beyond that available with stable beams alone. 
The radioactive secondary beams were produced  at the fragment 
separator FRS~\cite{frs92} by the fragmentation of primary $^{142}$Nd 
projectiles of about 900 MeV/nucleon in a thick beryllium  target. 
The FRS was set to select $^{124}$La and, in a second part of the experiment, 
also $^{107}$Sn projectiles which were then delivered to the ALADIN experimental setup.
All beams had a laboratory energy of 600 MeV/nucleon and were directed onto
reaction targets consisting of $^{\rm nat}$Sn with areal density 500 or 1000 mg/cm$^2$.

\begin{figure} [tbh]
\centerline{\includegraphics[width=8cm]{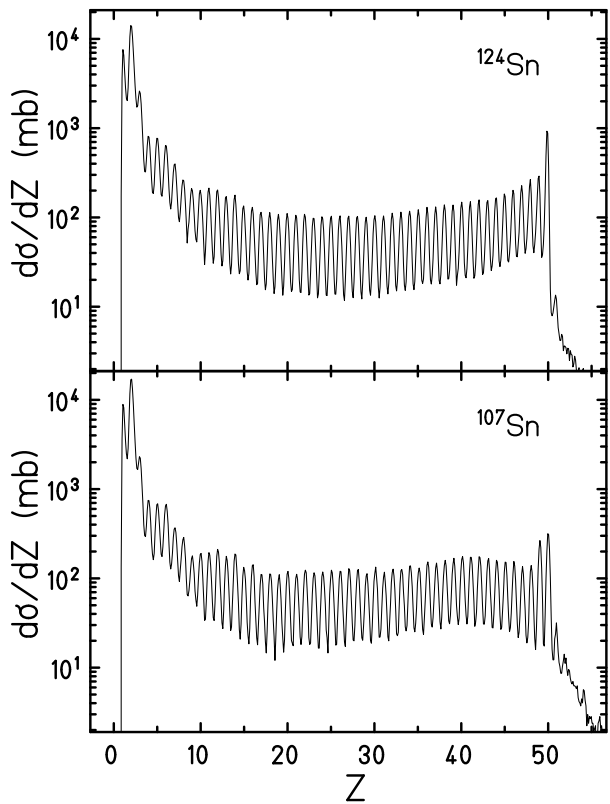}}
\caption{\small{Measured cross section as a function of the fragment atomic 
number $Z$ determined with the TP-MUSIC IV detector for $^{124}$Sn (top panel) and 
$^{107}$Sn projectiles (bottom panel). Note that the experimental trigger conditions
affected the yields for $Z >$~40. 
}}
\label{fig:z_exp}
\end{figure} 

For the runs with secondary beams, to reach the necessary beam intensity 
of about 10$^3$ particles/s with
the smallest possible mass-to-charge ratio $A/Z$, it was found necessary to
accept a distribution of neighboring nuclides together with the requested
$^{124}$La or $^{107}$Sn isotopes. 
The mean compositions of the nominal $^{124}$La ($^{107}$Sn) beams 
were $\langle Z \rangle$ = 56.8 (49.7) and $\langle A/Z \rangle$ = 2.19 (2.16), 
respectively~\cite{luk08}. 
Model studies confirm that these $\langle A/Z \rangle$ values are also representative for the 
spectator systems emerging after the initial stages of the reaction~\cite{LeFevre,Botvina02}.

\begin{figure} [htb]
\centerline{\includegraphics[width=8cm]{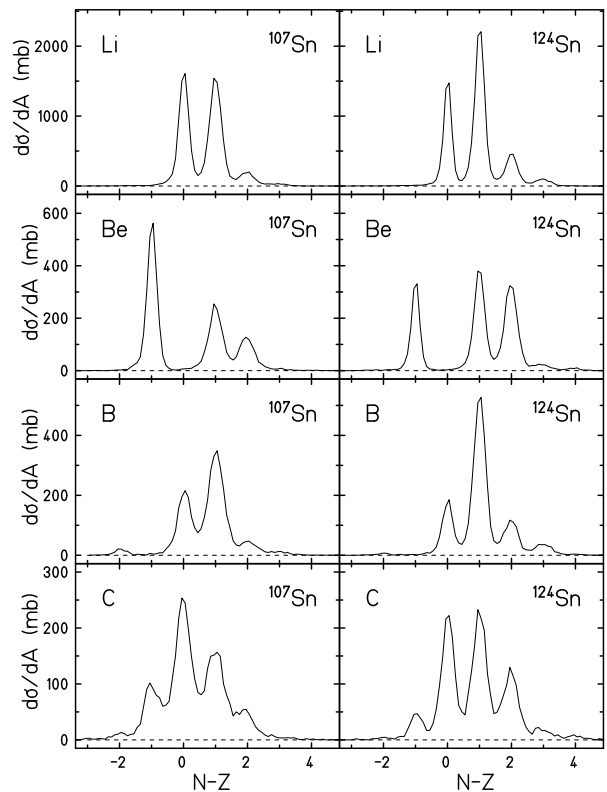}}
\caption{\small{Inclusive mass distributions for fragments with $3 \le Z \le6$ obtained for
$^{107}$Sn projectiles (left panels) and $^{124}$Sn projectiles (right panels). 
}}
\label{fig:a_exp}
\end{figure} 

The ALADIN experimental setup has been described previously~\cite{schuett96}. 
For the present experiment, it has been upgraded by constructing a new set of eight proportional 
readout chambers for the TP-MUSIC IV detector and by a redesign of all components of the 
electronic readout chain~\cite{sfienti02}. Digitization of the linear signals was performed 
with the aid of 40-MHz sampling ADC's right after the preamplifier stage, permitting digital 
filtering and background suppression. 
The readout planes on either side of the TP-MUSIC IV chamber were divided
into four proportional-counter sections and three ionization-anode sections. Each section extended 
over the full height of the chamber, and proportional and ionization sections alternated along 
the beam direction. With this arrangement, either type of detector could be used to record 
the fragment trajectories. Their horizontal coordinates, i.e. in the bending plane of the 
magnet, were obtained from the measured drift times while the vertical coordinates were 
determined from the positions along the proportional counters measured with a combination
of charge-division and pad-readout techniques. 

The threshold for fragment detection and identification was below atomic number $Z=2$, 
obtained with the proportional counters. The threshold of the ionization sections was higher
but they provided a superior $Z$ resolution for $Z>9$. Overall, the resolution 
$\Delta Z \leq 0.6$ (full width at half maximum) 
of the full detector was nearly independent of $Z$ up 
to the projectile $Z$ and permitted the individual identification of detected fragments 
according to their atomic number $Z$. 
Examples of recorded inclusive $Z$ spectra for the cases of stable $^{124}$Sn projectiles and 
radioactive $^{107}$Sn (nominal) projectiles are shown in Fig.~\ref{fig:z_exp}. They are given on 
an absolute scale but the recorded cross sections for very large fragments (about $Z > 40$) are
underestimated due to inefficiencies of the experimental trigger as discussed further below. 
The figure also shows that all elements are produced in multifragmentation reactions with an
intensity that varies rather smoothly with $Z$. For $^{124}$Sn, the recorded inclusive 
cross section for the production of fragments with $3 \le Z \le 49$ is 5.9 b. Charge 
pickup to $Z=51$ appears with 8 mb in the spectrum, a value fitting well into the
systematics collected for this process~\cite{rubehn96}. Very similar cross sections were obtained
for $^{107}$Sn projectiles. The most significant difference is the lower intensity near 
the atomic number of the projectile in the case of $^{107}$Sn, 
following from the reduced probability for 
neutron evaporation of neutron-poor heavy residue nuclei (see below). 
Traces of the neighboring nuclides in the nominal $^{107}$Sn beams are visible around $Z=50$.
Smaller differences
are observed in the range $Z<20$ where the odd-even $Z$ staggering is slightly more pronounced, 
on the level of 10\%, in the $^{107}$Sn case.

The magnetic rigidities of detected fragments were determined by backtracing their trajectories
through the magnetic field region of the ALADIN magnet, starting from the vectors determined 
with the TP-MUSIC IV detector. With an iterative procedure, the rigidity was varied until the 
calculated trajectory coincided with the measured coordinates of the projectile in the 
target plane, perpendicular to the beam direction.
A thin plastic detector of 200 $\mu$m in thickness, positioned close to the target and viewed by 
photomultipliers at four sides, was used for this purpose. 
The achieved position resolution amounted to $\approx 1.5$~mm, horizontally and vertically, 
over the active area of 4 x 4 cm$^2$ chosen for the secondary beam 
experiments.

The fragments, with velocities of approximately that of the projectile, can be assumed to be 
fully stripped of all electrons~\cite{begemann98}. Fragment momenta were determined from the 
measured rigidities and atomic numbers $Z$ and used to determine fragment masses by including 
the time-of-flight information given by the ALADIN Time-of-Flight Wall~\cite{schuett96}. 
The obtained time-of-flight resolution varied with $Z$, smoothly decreasing from $\approx 250$~ps 
(standard deviation) for lithium fragments to $\approx 100$~ps for fragments with $Z \ge 10$. 
The start time was derived from the projectile arrival time measured with a second thin plastic 
foil of 75 $\mu$m thickness, mounted upstream at an angle of 45$^{\circ}$ with 
respect to the beam direction and viewed by two photomultiplier tubes. The resolution obtained 
for the arrival time was $\approx 30$~ps (standard deviation).

\begin{figure} [htb]
\centerline{\includegraphics[width=8cm]{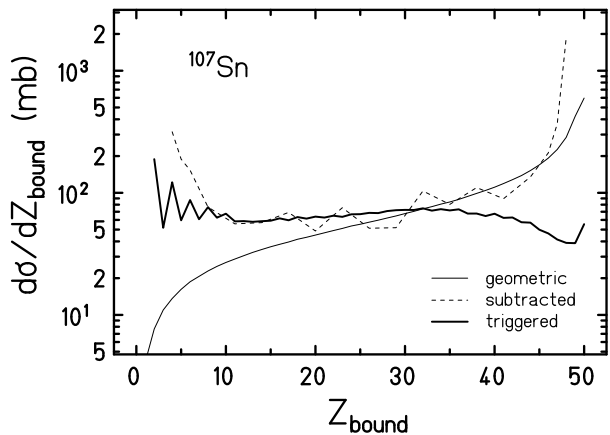}}
\caption{\small{Reaction cross section d$\sigma$/d$Z_{\rm bound}$ for $^{107}$Sn 
projectiles as recorded with the interaction trigger (thick solid line) in comparison with 
the expected value obtained with the subtraction technique (dashed line, see text).
The solid thin line represents, on the same abscissa, the cross section d$\sigma$/d$Z_{\rm spect}$
for the production of spectator nuclei with atomic number $Z_{\rm spect}$ according to the
geometric participant-spectator model of Ref.~\protect\cite{gosset77}.
}}
\label{fig:trig}
\end{figure} 
 
The obtained mass resolution is about 3\% for fragments with $Z \le 3$ (standard deviation), 
decreases to 1.5\% for $Z\geq 6$, and was found to be the same in the experiments
with stable and radioactive beams (Fig.~\ref{fig:a_exp}). Masses are individually resolved for 
fragments with atomic number $Z \leq 10$. For optimum resolution, correct timing signals from 
both the front and the back layer of the time-of-flight wall had to be required. Background-free 
mass spectra, without spurious peaks due to misidentified charges, were obtained by requiring
that the $Z$ values delivered by the TP-MUSIC IV and the two layers of the time-of-flight 
wall all coincided within $\pm 0.5$~units of the adopted integer value. Because of the dead 
zones between adjacent scintillator slats and the reduced charge resolution of the
time-of-flight wall for larger $Z$, this condition was met by one-half to one-fourth of the
fragments with $Z \le 10$. The cross section distributions shown in 
Fig.~\ref{fig:a_exp} are, therefore, normalized with respect to the recorded $Z$ yields. 
The significant differences of the isotope yields for neutron-poor and neutron-rich systems 
are evident. They are largest for the weaker but still identifiable peaks such as, e.g., $^8$B 
on the neutron-poor or $^{12}$Be and $^{16}$C on the neutron-rich side of the spectra.
Isotopically resolved production cross sections are shown and discussed in 
Sec.~\ref{sec:iso}.

The acceptance of the ALADIN forward spectrometer, in the geometry of the present experiment,
was $\pm 10.2^{\circ}$ for $N=Z$ fragments with beam velocity in the horizontal direction,
i.e. in the bending plane, and $\pm 4.5^{\circ}$ in vertical direction. 
At the beam energy of 600 MeV, this corresponded to an acceptance increasing with $Z$ from 
about 90\% for projectile fragments with $Z=3$ to values exceeding 95\% for $Z \ge 6$. 
These values were obtained by fitting the measured momentum distributions in transverse directions
with Gaussian functions and by estimating the yields lost in the tails. 
Very similar acceptances have been observed in earlier experiments at the same beam 
energy~\cite{schuett96}. 

Because of the selective coverage of the projectile-spectator decay,
the quantity $Z_{\rm bound}$ defined as the sum of the atomic numbers $Z_i$ of all detected
fragments with $Z_i \geq$ 2 has been chosen as the principal variable for event sorting. 
$Z_{\rm bound}$ represents approximately the charge of the primary spectator system, 
apart from emitted hydrogen isotopes, and is monotonically correlated with the impact 
parameter of the reaction \cite{ogilvie91}. 
The excitation energy per nucleon of the spectator system is 
inversely correlated with $Z_{\rm bound}$~\cite{schuett96}.

Trigger signals for reactions in the target were derived from four plastic-scintillator paddles,
each 5 mm thick, of 8 x 50 cm$^2$ active area, and viewed by photomultiplier tubes at both 
ends. They were positioned approximately 50 cm downstream from the target 
at angles outside the acceptance of 
the spectrometer. The condition that at least one of them fired was met with nearly 100\% 
probability by events with moderate to large charged-particle multiplicities. This is 
demonstrated in Fig.~\ref{fig:trig} in which the rate of interaction triggers obtained 
in this way is compared to the expected event rate. The latter was obtained by subtracting 
event rates recorded with and without a reaction target from each other. 

The most notable deviation from the expected rate is seen at large $Z_{\rm bound}$ at which 
the expected yields increase rapidly while the cross section recorded with the interaction
trigger remains fairly flat as a function of $Z_{\rm bound}$. A rise of the cross section 
is expected on the basis of the reaction geometry as shown in the figure. The smooth thin 
line represents the cross section predicted by the spherical participant-spectator 
model~\cite{gosset77} for $^{107}$Sn as a function of the spectator charge $Z_{\rm spect}$,
plotted on the same $Z$ scale. At small $Z_{\rm bound}$, the geometric prediction 
falls considerably below the measured yields because emitted hydrogen isotopes are not counted 
in $Z_{\rm bound}$, but probably also because the sharp-edge spherical geometry is less 
realistic for the more central collisions.

\begin{figure} [tbh]
\centerline{\includegraphics[width=8cm]{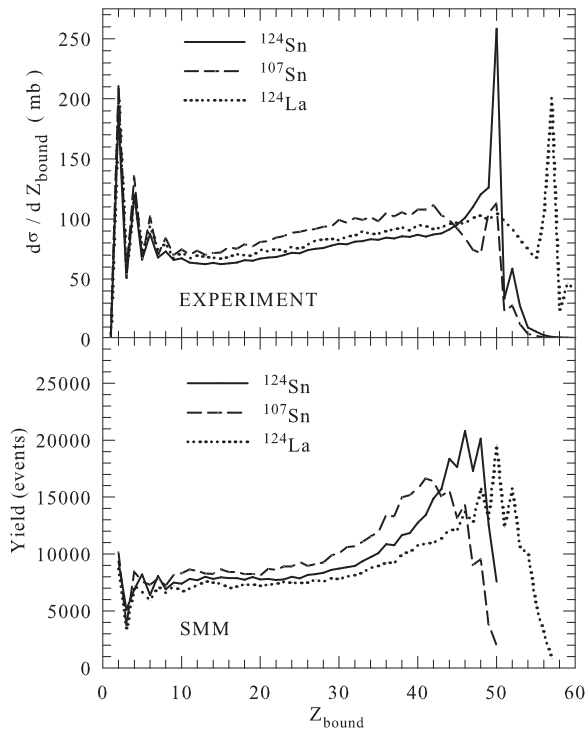}}
\caption{\small{Top panel: Triggered reaction cross section as a function of
$Z_{\rm bound}$ for the three projectiles as indicated.
Bottom panel: The corresponding distributions as obtained from ensemble calculations 
for 500000 reaction events with standard parameters (see text).
}}
\label{fig:zbound}
\end{figure}

The cross sections d$\sigma$/d$Z_{\rm bound}$ recorded in this way for the three cases of
$^{107}$Sn, $^{124}$Sn, and $^{124}$La projectiles are shown in Fig.~\ref{fig:zbound} on a
linear scale. Their
integrated values of 4.3 b ($Z_{\rm bound} \le 48$), 3.9 b ($Z_{\rm bound} \le 49$), 
and 4.6 b ($Z_{\rm bound} \le 55$), respectively, compare well with a calculated reaction 
cross section $\sigma_R = 4.8$~b ($r_0 = 1.25$~fm) for Sn + Sn collisions. It confirms
that the major part of the cross section is recorded with high probability, as shown 
in Fig.~\ref{fig:trig}. 
Only for very peripheral and extremely central events does a bias exist against events with small
charged-particle multiplicity.   

Overall, the cross sections vary rather smoothly with $Z_{\rm bound}$. The drop of the
cross section toward the projectile charge $Z_0$ in the neutron-poor cases, while it rises 
for $^{124}$Sn, has already been noted in the $Z$ spectra (Fig.~\ref{fig:z_exp}) but is 
observed more clearly on the linear scale (Fig.~\ref{fig:zbound}). 
The fluctuations at small $Z_{\rm bound}$ are caused by multi-alpha events which 
enhance the even channels. 
The high event rates for $Z_{\rm bound} =Z_0$ are most likely due to $\delta$ electrons 
causing occasional spurious triggers of non-interacting projectiles.

\section{STATISTICAL APPROACH TO MULTIFRAGMENTATION}
\label{sec:sta}

\subsection{Reaction scenario and strategy}

In the scenario assumed for the statistical analysis, the multifragmentation process is 
subdivided into (1) a dynamical stage leading to the
formation of an equilibrated nuclear system, (2) the disassembly of the system into 
individual primary fragments, and (3) the subsequent de-excitation of the hot primary fragments.
For the description of the nonequilibrium stage of peripheral heavy-ion collisions,
a variety of models can be used. Realistic calculations of ensembles of excited residual nuclei 
that undergo multifragmentation were first performed with the intranuclear cascade (INC) 
model~\cite{Botvina94}. 
More recent examples of hybrid calculations 
using INC or Boltzmann-Uehling-Uhlenbeck (BUU) models for the dynamical 
stage have been reported in Refs.~\cite{legentil08,gait09}. 

The dynamical models agree in that the character of the dynamical
evolution changes after a few rescatterings of incident nucleons, when high
energy particles ("participants") leave the system.
However, the time needed for equilibration and transition to the
statistical description is uncertain and model dependent~\cite{Barz,raduta07}.
It is estimated to be around 100 fm/$c$ for spectator matter 
but the required parameters, i.e., the excitation energies, mass numbers, and charges 
of the predicted equilibrated sources, vary significantly even over longer times.
To evade these ambiguities and other uncertainties related to the coupling with a particular 
dynamical model, alternative strategies have been 
developed~\cite{xi97,Barz,Botvina92,Gross,Botvina95,deses96,Raduta,lefevre04}
and are also followed here. 
With the results of dynamical simulations as qualitative guidelines, 
the exact parameters of the thermalized sources are obtained from the analysis of the
experimental fragmentation data. Usually, only a subgroup of observables is required for the
determination of the source parameters (see, e.g., Refs.~\cite{xi97,lefevre04}). 
The success of the statistical description is then judged from the quality of predictions 
for other observables and for the correlations between them. 

\subsection{Ensemble parameters}

\begin{figure} [tbh]
\centerline{\includegraphics[width=8cm]{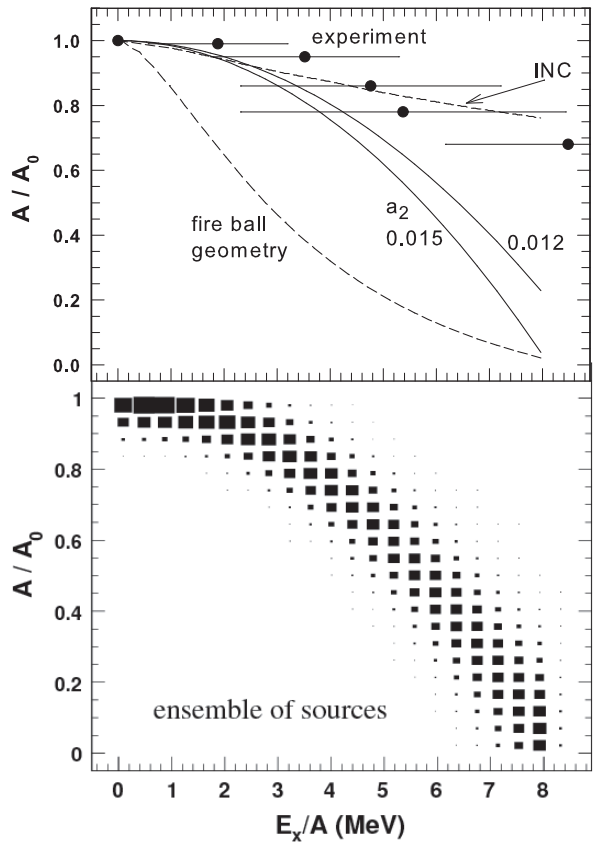}}
\caption{\small{Top panel: Mean reduced mass number $A/A_0$ of relativistic 
projectile residuals as a function of their excitation energy $E_x/A$ 
according to the fireball model of Gosset {\it et al.}~\protect\cite{gosset77}, 
calculated for $^{107}$Sn projectiles, 
and Intra-Nuclear Cascade calculations for $^{197}$Au+C~\protect\cite{Botvina94} 
(dashed lines), according to the experimental results for $^{197}$Au nuclei 
reported by Pochodzalla {\it et al.}~\protect\cite{poch95} 
and including the widths in $E_x/A$ (dots), and according to the 
parametrization used in this work for $a_2=0.012$ and 0.015 MeV$^{-2}$ (solid lines). 
Bottom panel: Ensemble of hot thermal sources represented in
a scatter plot of reduced mass number $A/A_0$ versus 
excitation energy $E_x/A$, as used in the
SMM calculations. The frequency of the individual sources is proportional
to the area of the squares.
}}
\label{fig:eoa}
\end{figure}

The correlations between mass and excitation energy obtained from an INC calculation~\cite{Botvina94}
and of the geometric participant-spectator model~\cite{gosset77} are shown in 
the top panel of Fig.~\ref{fig:eoa} together with the experimental result 
for $^{197}$Au + $^{197}$Au at 600 MeV/nucleon~\cite{poch95}. 
The general correlation of an increasing excitation energy per nucleon with a decreasing mass 
of the produced projectile spectators is a feature common to all cases
but the values differ considerably. The increase in surface energy, solely considered in the
geometric model, is much smaller than the transferred energies predicted by the INC or obtained from
the experiment. The latter two results are fairly consistent with each other but the excitation
energies were found to be larger than what is needed for satisfactory statistical 
descriptions~\cite{schuett96,Barz,Botvina92,Gross}. A considerable part of the accumulated energy 
is not distributed statistically at freeze-out, as evident from the fragment kinetic energies which 
are much higher than their thermal expectations~\cite{odeh00}. 

In the present work, we follow the procedure used by Botvina {\it et al.}~\cite{Botvina95} 
which was rather effective in describing the multifragmentation of relativistic $^{197}$Au 
projectiles, including their correlations and dispersions. 
There, the average masses of the equilibrated sources $A_s$ were parameterized as
$A_s/A_0 =1-a_1(E_x/A_s)-a_2(E_x/A_s)^2$, where
$E_x$ is the excitation energy of the sources in MeV, and $A_0$ is the
projectile mass. The correlations obtained with parameters $a_1=0.001$~MeV$^{-1}$ 
and $a_2=0.015$~MeV$^{-2}$, used in Ref.~\cite{Botvina95}, as well as $a_2=0.012$~MeV$^{-2}$ are shown in 
Fig.~\ref{fig:eoa}. It is interesting to note that a nearly identical correlation was obtained 
by D\'{e}sesquelles {\it et al.}~\cite{deses96} with a linear backtracing algorithm based on the 
SMM and on the fragmentation observables measured for $^{197}$Au + Cu at 600 MeV/nucleon. 
As argued in Ref.~\cite{Barz}, a dependence of this kind can be considered as consistent 
with BUU calculations. 

The ensemble of thermal sources adopted for the present analysis, for $a_2=$0.015 MeV$^{-2}$,
 is given in the 
bottom panel of Fig.~\ref{fig:eoa} as a weighted distribution in the $A_s/A_0$ 
versus $E^*/A_s$ plane. The widths are empirically adjusted, guided by the earlier 
results for $^{197}$Au fragmentation~\cite{Botvina95}. 
The yield distributions as functions of the reduced mass $A/A_0$ and excitation energy per
nucleon were also generated empirically, in a similar manner as described in the same reference. 
Identical parameters were used for the three reaction systems. 
The charges of the sources were determined by assuming that the charge-to-mass ratio $Z/A$
remains that of the initial projectiles. This is expected for the initial reaction phase, as
long as nucleon-nucleon dynamics dominates, and supported by INC and relativistic BUU 
calculations for similar collision systems~\cite{LeFevre,Botvina02}. 
Further arguments in favor of this assumption derived from the experimental 
neutron-to-proton ratios of emitted fragments are discussed in Sec.~\ref{sec:dis}.

\subsection{Model description}
As the model for multifragmentation, describing the reaction stages (2) and (3) defined above,
the statistical multifragmentation model (SMM, for a review see Ref.~\cite{SMM}) is used.
The SMM assumes statistical equilibrium of the excited nuclear system with
mass number $A_s$, charge $Z_s$, and excitation energy $E_x$ (above the ground
state) within a low-density freeze-out volume.
All breakup channels (partitions $\{p\}$)
composed of nucleons and excited fragments are considered and the
conservation of baryon number, electric charge, and energy are taken into account.
Besides the breakup channels, also the compound-nucleus channels are included 
and the competition between all channels is permitted.
In this way, the SMM covers the conventional evaporation and fission processes occurring
at low excitation energy as well as the transition region between the low- and high-energy 
de-excitation regimes. In the thermodynamic limit, as demonstrated 
in Refs.~\cite{gupta98,bugaev01,chaud09},
the SMM is consistent with the nuclear liquid-gas phase transition when the liquid phase is 
represented by an infinite nuclear cluster.

In the model, light nuclei with mass number $A \leq 4$ and charge
$Z \leq 2$ are treated as elementary stable particles with masses and
spins taken from the nuclear tables ("nuclear gas"). Only translational
degrees of freedom of these particles contribute to the entropy of the
system. Fragments  with $A > 4$ are treated as heated nuclear liquid drops. 
Their individual free energies $F_{AZ}$ are parametrized as a sum of the bulk,
surface, Coulomb and symmetry energy contributions
\begin{equation}
F_{AZ}=F^{B}_{AZ}+F^{S}_{AZ}+E^{C}_{AZ}+E^{sym}_{AZ}.
\end{equation}
The standard expressions for these terms are
$F^{B}_{AZ}=(-W_0-T^2/\epsilon_0)A$, where $T$ is the temperature,
the parameter $\epsilon_0$ is related to the level density, and
$W_0 = 16$~MeV is the binding energy of infinite nuclear matter;
$F^{S}_{AZ}=B_0A^{2/3}((T^2_c-T^2)/(T^2_c+T^2))^{5/4}$, where
$B_0=18$~MeV is the surface energy coefficient and $T_c=18$~MeV is the critical
temperature of infinite nuclear matter; $E^{C}_{AZ}=cZ^2/A^{1/3}$, where
$c=(3/5)(e^2/r_0)(1-(\rho/\rho_0)^{1/3})$ is the Coulomb parameter (obtained
in the Wigner-Seitz approximation) with the charge unit $e$ and
$r_0=$~1.17 fm;  $E^{sym}_{AZ}=\gamma (A-2Z)^2/A$, where
$\gamma = 25$~MeV is the symmetry energy parameter.
These parameters are those of the Bethe-Weizs\"acker formula and correspond
to the assumption of isolated fragments with normal density in the
freeze-out configuration, an assumption found to be quite successful in
many applications.

Within the microcanonical treatment \cite{SMM,Botvina01}
the statistical weight of a partition $p$ is calculated as
\begin{eqnarray}
W_{p} \propto exp~S_{p},
\end{eqnarray}
where $S_{p}$ is the corresponding entropy, depending on the fragments
in this partition as well as on the parameters of the system.
In the grand canonical treatment of the SMM \cite{Botvina85},
after integrating out translational degrees of freedom, 
the mean multiplicity of nuclear fragments with $A$ and $Z$
can be written as
\begin{eqnarray}
\label{naz} \langle N_{AZ} \rangle =
g_{AZ}\frac{V_{f}}{\lambda_T^3}A^{3/2} {\rm
exp}\left[-\frac{1}{T}\left(F_{AZ}-\mu A-\nu
Z\right)\right]. 
\end{eqnarray}
Here $g_{AZ}$ is the ground-state degeneracy factor of species
$(A,Z)$, $\lambda_T=\left(2\pi\hbar^2/m_NT\right)^{1/2}$ is the
nucleon thermal wavelength, and $m_N \approx 939$ MeV/$c^2$ is the
average nucleon mass. $V_f$ is the free volume available for the
translational motion of fragments. 
The free energy $F_{AZ}=F_{AZ}(T,\rho)$ of fragments with $A>4$ and $Z>2$, parametrized
as given above, is a function of the temperature and density.
The chemical potentials $\mu$ and $\nu$ are found from the mass and charge constraints:
\begin{equation} \label{eq:ma2}
\sum_{(A,Z)}\langle N_{AZ}\rangle A=A_{s},~~ \sum_{(A,Z)}\langle
N_{AZ}\rangle Z=Z_{s}. 
\end{equation}
For the freeze-out density, one-third the normal nuclear density is assumed.  
This is a standard value, used previously in many successful applications 
and consistent with independent experimental determinations
of the freeze-out density~\cite{fritz99,viola04}. 

For generating individual multifragmentation events, the methods proposed in 
Refs.~\cite{SMM,Botvina95} have been used. 
Individual sources are randomly selected from the ensemble of excited sources  
according to their probability of occurrence within the ensemble (Fig.~\ref{fig:eoa}) 
up to a total of typically $10^{5}$ to $10^{6}$ sources. 
Their breakup is calculated with the SMM by the Monte-Carlo method described  
in Ref.~\cite{SMM} which has the advantage of a high efficiency.
In particular, at low excitation energies, when partitions of primary hot fragments with small
total multiplicities $M$ dominate, the probabilities of all partitions with $M \leq 3$, 
including the compound nucleus formation, are included. 

As a first step, partitions are randomly selected from the ensemble of all possible partitions
and their probabilities are determined. For small fragment multiplicities, the microcanonical
sampling according to Eq.~[2] can be performed. 
At higher excitation energy, typically above $E_x \approx 4$~MeV/nucleon, 
the masses and charges of 
primary hot fragments are generated according to the grand-canonical distribution (Eq.~[3]), 
however, by requiring the exact mass number and charge conservations in each partition. 
The temperature of a partition is found iteratively by adjusting it so that the total energy 
is conserved. It is, therefore, referred to as the microcanonical temperature of the partition.
After their formation in the freeze-out volume, the fragments propagate
independently in their mutual Coulomb fields and undergo secondary
decays. The deexcitation of the hot primary fragments proceeds via
evaporation, fission, or via Fermi breakup for primary fragments with
$A \le 16$~\cite{Botvina87}. 

Because the calculated events contain all particles and fragments with their masses, charges, 
and momenta including those produced in secondary decays, 
they can be treated in the same way as experimental events. Sorting into event groups can be
performed and group averages and distributions or correlations can be determined.
Before comparisons with experimental data are made, the resulting partitions of cold fragments 
are subjected to the ALADIN experimental filter by randomly removing fragments according 
to the $Z$-dependent acceptance given in Sec.~\ref{sec:exp}. 

\begin{figure} [htb]
\centerline{\includegraphics[width=8.5cm]{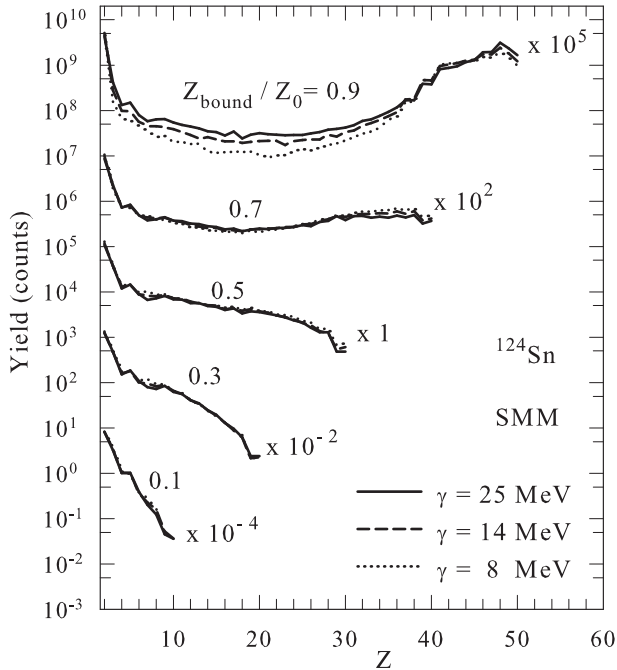}}
\caption{\small{
Relative yields of fragments for 10-unit-wide intervals of $Z_{\rm bound}$ 
as obtained with the SMM for the standard symmetry-term coefficient $\gamma = 25$~MeV (solid lines)
and with reduced coefficients $\gamma = 14$~and 8~MeV (dashed and dotted lines, respectively)
as a function of the fragment $Z$ for the case of $^{124}$Sn projectiles. 
The centers of the five $Z_{\rm bound}$ intervals relative to $Z_0$ 
and the scale factors used for displaying
the yields are indicated in the figure.   
}}
\label{fig:gonz}
\end{figure} 

\subsection {In-medium modification of fragment properties}

The liquid-drop model for the description of the emerging fragments is an
essential and very successful ingredient of the SMM~\cite{SMM}. 
It is based on the concept of an idealized chemical freeze-out,  
defined as the 
moment in time at which the nuclear forces have ceased to act between 
fragments. Their action within fragments can then be summarized with a liquid-drop
description of the fragment properties with standard parameters. 
Only the Coulomb energy is reduced due to the 
presence of other fragments, an effect reasonably well accounted for within
the Wigner-Seitz approximation~\cite{SMM}.
However, in a more realistic treatment of the breakup stage, 
the excited primary fragments may have to be considered as not only excited but also having 
modified properties due to residual interactions with the environment. 
Their shapes and average densities may deviate from the equilibrium properties 
of isolated nuclei~\cite{Larionov,baran02}. Evidence for such effects has already been obtained
from dynamical simulations~\cite{ono04} and from dynamical or statistical analyses of 
isotopic effects in a variety of reactions~\cite{LeFevre,Botvina05,iglio06,souliotis07}.
Within the statistical approach, they can be accounted for in the fragment
free energies by changing the corresponding liquid-drop parameters.
In the following, we search for such effects by testing the sensitivity of fragment 
observables to the surface and symmetry energy terms in
comparison with the present data. 

At the last stage of the multifragmentation process, the hot primary
fragments undergo deexcitation and propagate in the mutual Coulomb
field. The realistic description of this stage is essential for 
obtaining reliable final fragment yields. To the extent that this can be achieved, 
the hot breakup stage will become accessible, which is the preeminent aim of this work. 
Possible modifications of the
fragment properties must be taken into account in the first 
deexcitation steps, as the hot fragments are still surrounded by other species, 
while, at the end of the evaporation cascade, the standard properties
must be restored. In the actual calculations, a linear interpolation between 
these two limiting cases has been introduced 
in the interval
of excitation energies between zero and $E_x^{\rm int} = 1$~MeV/nucleon~\cite{buyuk05}. 
Energy and momentum conservation are observed throughout this process.

Possible in-medium modifications of light clusters with $A \le 4$, treated as elementary 
particles in the SMM, are not considered here. This seems justified because of 
their small primary multiplicities in the multifragmentation channels. For 
example, in the primary partitions of a single source $^{124}$Sn at $E_x = 5$~MeV/nucleon, 
$^{2,3}$H and $^{3,4}$He nuclei appear with mean multiplicities of 0.14, 0.07, 0.03, 
and 0.29, respectively. Together they amount to a fraction of less than 1.5\% of the total 
mass of the system. 
The majority of light clusters observed in the final partitions stem from secondary 
decays~\cite{buyuk08}.
Modifications of light clusters have been intensively studied with microscopic approaches,
preferentially however in temperature-density domains at which larger fragments are of 
reduced importance (see, e.g., Refs.~\cite{typel10,nato10} and references given therein).

\section{Charge characteristics of fragment partitions}
\label{sec:cha}

\subsection{Ensemble normalization}

For event characterization, the total charge bound in fragments with $Z\geq2$, 
$Z_{\rm bound}$, has been introduced as in previous reports on ALADIN results (see, e.g.,
Refs.~\cite{xi97,kreutz,schuett96,Botvina95}).
Small $Z_{\rm bound}$ values correspond to high excitation energies
of the sources that disintegrate predominantly into very
light clusters ("fall" of multifragmentation). Large values of
$Z_{\rm bound}$ correspond to low excitation energies, at which the
decay changes its character from evaporationlike/fissionlike 
processes to multifragmentation ("rise" of multifragmentation).

\begin{figure} [htb]
\centerline{\includegraphics[width=7.5cm]{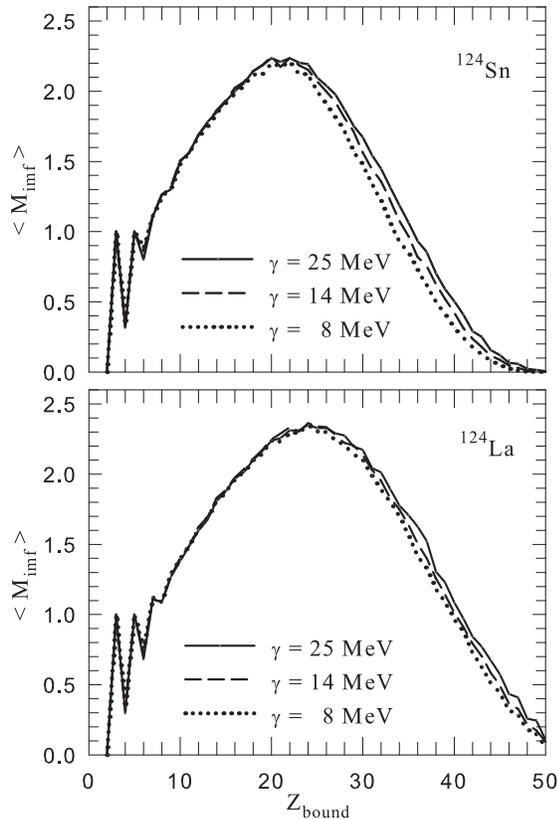}}
\caption{\small{Mean multiplicity of intermediate-mass fragments, 
$\langle M_{\rm imf} \rangle$,
as a function of $Z_{\rm bound}$ as obtained with the SMM for the standard symmetry-term 
coefficient $\gamma = 25$~MeV (solid lines) and with reduced coefficients $\gamma = 14$~and 8 MeV 
(dashed and dotted lines, respectively), for reactions of $^{124}$Sn (top) and $^{124}$La 
(bottom) projectiles.}}
\label{fig:gonimf}
\end{figure} 

In Fig.~\ref{fig:zbound}, besides the experimental also the theoretical event 
distributions as a function of $Z_{\rm bound}$ are shown for the three 
projectiles under investigation, $^{107}$Sn, $^{124}$Sn, and $^{124}$La. 
The theoretical results are obtained with the ensembles introduced in the previous section
(Fig.~\ref{fig:eoa}) and with standard assumptions regarding the fragment properties,
i.e. $B_0 = 18$~MeV and $\gamma = 25$~MeV for the surface- and symmetry-term coefficients.
As $Z_{\rm bound}$ is approximately equal to the charge of the spectator system and only 
slightly affected by its excitation energy, the comparison is mainly a test of how well the 
ensemble parameters are chosen. Modifications of the liquid-drop parameters have a very small 
effect on the $Z_{\rm bound}$ spectra.

\begin{figure} [tbh]
\centerline{\includegraphics[width=8cm]{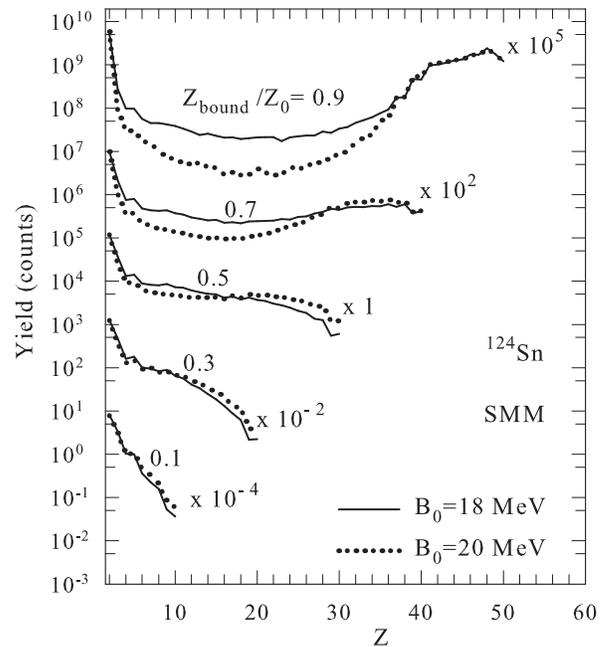}}
\caption{\small{Relative yields of fragments for 10-unit-wide intervals of $Z_{\rm bound}$ 
as obtained with the SMM for the standard surface-term coefficient $B_0 = 18$~MeV (solid lines) 
and with the modified coefficient $B_0 = 20$~MeV (dotted line)
as a function of the fragment $Z$ for the case of $^{124}$Sn projectiles. 
The centers of the five $Z_{\rm bound}$ intervals and the scale factors used for displaying
the relative yields are indicated in the figure.}}
\label{fig:bonz}
\end{figure} 

The agreement between theory and experiment is overall very satisfactory and sufficient for 
performing the following analyses in individual intervals of $Z_{\rm bound}$. 
A larger disagreement 
is observed in the region of $Z_{\rm bound}> 40$ for the Sn projectiles and 
at $Z_{\rm bound} > 46$ for the $^{124}$La projectile (Fig.~\ref{fig:zbound}).
As discussed above, it may be partly of experimental origin.
The discrepancies at very small $Z_{\rm bound} <10$ are mainly occurring in channels 
dominated by multi-$\alpha$ events. Here, the larger experimental yields may come from
more central collisions 
that are beyond the statistical treatment of the spectator multifragmentation studied here.
For these reasons, the following analysis mainly focuses on the intermediate
range of $Z_{\rm bound} \approx$ 10--40 (46 in the case of $^{124}$La), constituting 
the major part of the rise and fall of nuclear multifragmentation. 

For the quantitative comparison of theory and experiment, the SMM ensemble calculations 
were globally normalized with respect to the measured $Z_{\rm bound}$ cross sections in the 
interval $10 <Z_{\rm bound} \le30$ for which the agreement is best. The obtained factors
are 0.00937 mb, 0.00804 mb, and 0.0106 mb per theoretical event for $^{107}$Sn, $^{124}$Sn, 
and $^{124}$La projectiles, respectively, i.e on average about 100 events per mb were calculated
with the SMM.

\begin{figure} [tbh]
\centerline{\includegraphics[width=7.3cm]{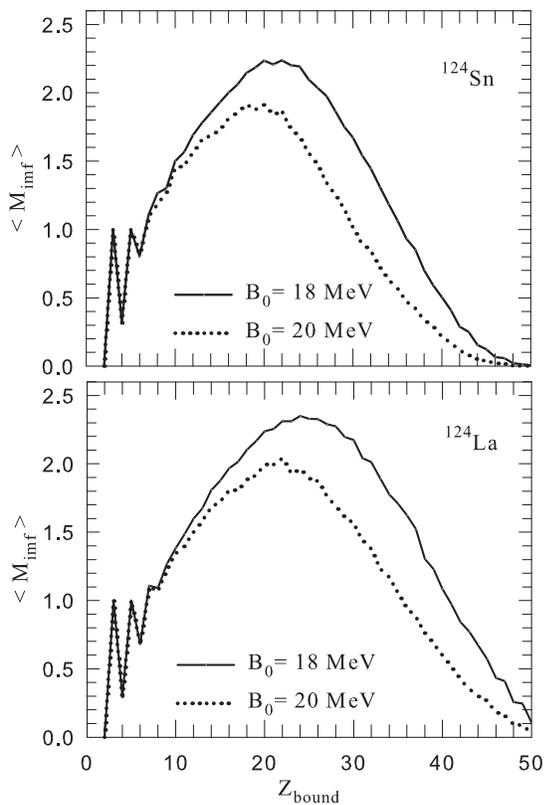}}
\caption{\small{
Mean multiplicity of intermediate-mass fragments, $\langle M_{\rm imf} \rangle$,
as a function of $Z_{\rm bound}$ obtained with the SMM for the standard surface-term 
coefficient $B_0 = 18$~MeV (solid lines) and with the modified coefficient $B_0 = 20$~MeV 
(dotted lines), for reactions of $^{124}$Sn (top) and $^{124}$La (bottom) projectiles.}}
\label{fig:bonimf}
\end{figure} 

\subsection{Sensitivity to liquid-drop parameters} 

As a first step, the sensitivity of the fragment-charge distributions and correlations
to the intrinsic properties of these fragments was investigated with the SMM.
Ensemble calculations were performed for different parameters of the liquid-drop
description of the produced fragments and analyzed as a function of the obtained
$Z_{\rm bound}$. To permit a more detailed comparison and to include the case of $^{124}$La
on the same scale, the reduced $Z_{\rm bound}/Z_0$, normalized with respect to the atomic number 
$Z_0$ of the projectile, has been used for some observables and 
subdivided into five bins (0--0.2, 0.2--0.4, etc.), labeled by their central values
(0.1, 0.3, etc.) in the legends of the figures.

For the case of $^{124}$Sn, the relative fragment yields as a function of $Z$ sorted into these
five bins are shown in Fig.~\ref{fig:gonz}. The charge distribution evolves from a so-called
"U-shaped" distribution, with domination of evaporation of the compound
nucleus and asymmetric binary decay during the onset of multifragmentation,
to a rapidly dropping exponential distribution. 
This evolution is a characteristic feature of the present type of spectator fragmentation
\cite{kreutz,ogilvie91,hauger00}. 
The variation of the symmetry-term coefficient in the calculations, 
down to one third of its standard value, has a negligible effect on the resulting charge
distributions except in the bin of largest $Z_{\rm bound}$. There the fragment yields in the
intermediate-mass range which are already small are further reduced if smaller $\gamma$ 
parameters are chosen. 

\begin{figure} [htb]
\centerline{\includegraphics[width=8cm]{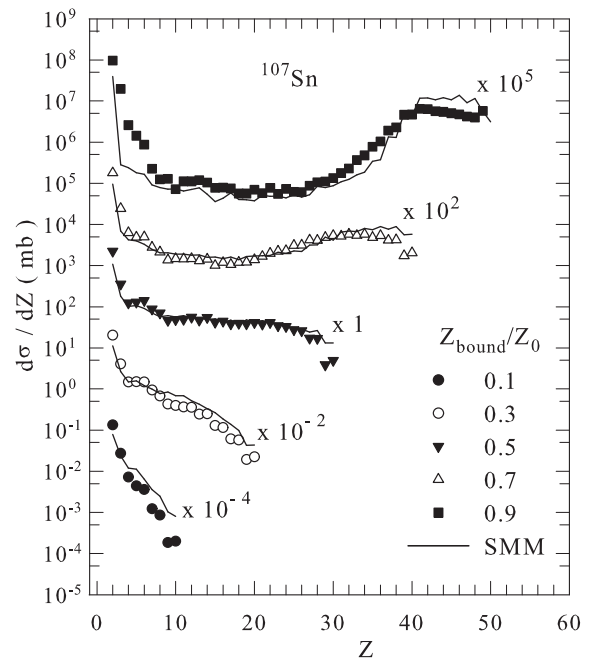}}
\caption{\small{Experimental cross sections d$\sigma$/d$Z$ for the fragment production following
collisions of $^{107}$Sn projectiles sorted into five intervals of $Z_{\rm bound}/Z_0$ with centers
as indicated and width 0.2 (symbols) in comparison with normalized SMM calculations (lines). 
Different scale factors were used for displaying the cross sections as indicated.
}}
\label{fig:z107}
\end{figure} 

The same conclusion can be drawn from the results for the multiplicities of
intermediate-mass fragments (IMF, $3 \le Z \le 20$) shown in Fig.~\ref{fig:gonimf}
as functions of $Z_{\rm bound}$ for the neutron-rich $^{124}$Sn and the neutron-poor 
$^{124}$La projectiles. The effect of varying the symmetry-term coefficient is very small
and visible only at the rise of multifragmentation on the right half of the figure.
The fall part is expected to be universal because of the auto-correlation of the two 
quantities $M_{\rm imf}$ and $Z_{\rm bound}$~\cite{hubele92}.
The autocorrelations dominate in the cases of $Z_{\rm bound}$~=~3 and 5 which can only be reached
with partitions containing exactly one lithium fragment.

The surface energy of fragments is important for the
charge yields \cite{Botvina06}. As demonstrated in Figs.~\ref{fig:bonz} and \ref{fig:bonimf},
the relatively small variation of the surface-energy coefficient $B_0$ from 18 to 20 MeV
leads to considerable changes in the fragment production. The larger surface energy
suppresses multifragmentation, leading to smaller IMF yields and multiplicities. 
And, vice versa, decreasing $B_0$ results in an intensive breakup of
nuclei already at low excitation energy and to steeper $Z$ spectra. 
Because of the mentioned autocorrelations, the influence of $B_0$ decreases with
decreasing $Z_{\rm bound}$, as evident from both figures.

\begin{figure} [htb]
\centerline{\includegraphics[width=8cm]{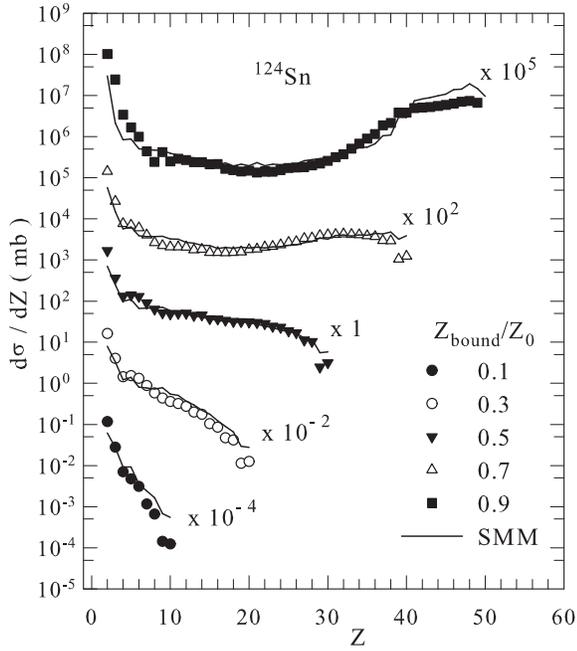}}
\caption{\small{As Fig.~\protect\ref{fig:z107} but for the case of $^{124}$Sn
projectiles.
}}
\label{fig:z124}
\end{figure} 

\begin{figure} [htb]
\centerline{\includegraphics[width=8cm]{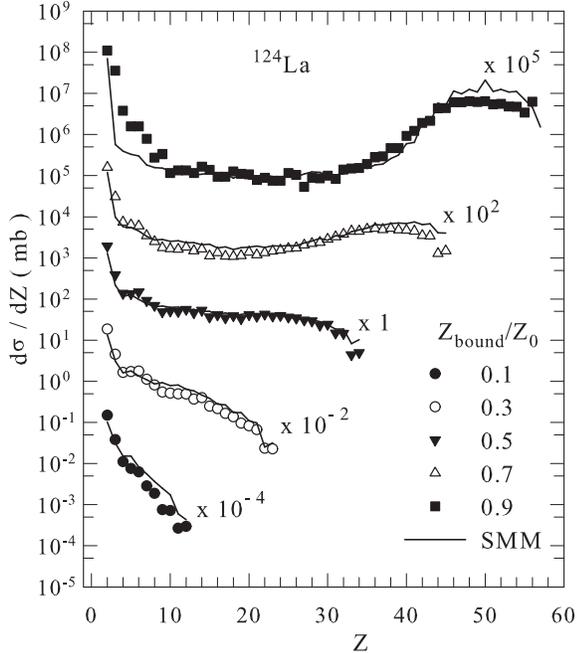}}
\caption{\small{As Fig.~\protect\ref{fig:z107} but for the case of $^{124}$La
projectiles.
}}
\label{fig:zlan}
\end{figure} 

\subsection{Data comparison}

The measured cross sections d$\sigma$/d$Z$ for fragment production in reactions initiated
by $^{107}$Sn, $^{124}$Sn, and $^{124}$La projectiles are shown in 
Figs.~\ref{fig:z107}--\ref{fig:zlan}. The data have been sorted into five bins of the
reduced bound charge $Z_{\rm bound}/Z_0$ of width 0.2 spanning the range up to 
$Z_{\rm bound} = Z_0$. The evolution from U-shaped to exponential
$Z$ spectra, already discussed in the previous section, appears without noticeable 
differences between the three cases.

\begin{figure} [htb]
\centerline{\includegraphics[width=7.5cm]{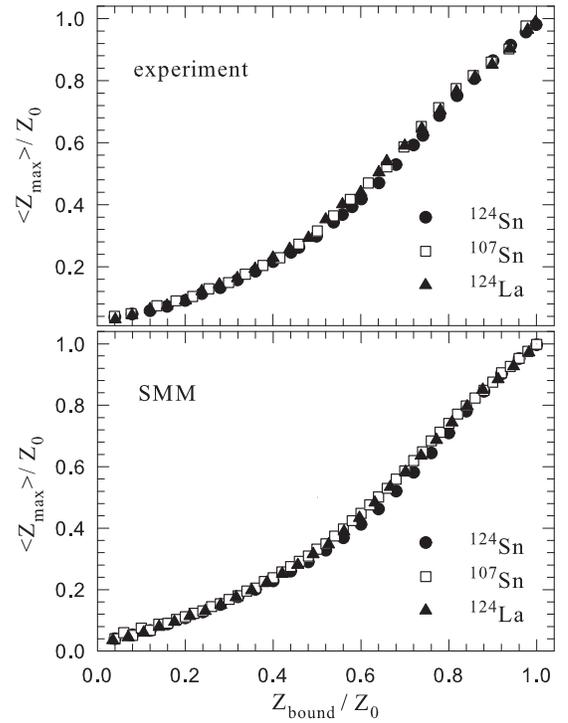}}
\caption{\small{Experimental results (top panel) and results of SMM calculations (bottom panel)
for the mean maximum charge $\langle Z_{\rm max} \rangle$ within an event versus $Z_{\rm bound}$, 
both normalized with respect to $Z_0$, for the three studied projectiles.  
}}
\label{fig:zmaxsmm}
\end{figure} 

\begin{figure} [htb]
\centerline{\includegraphics[width=7cm]{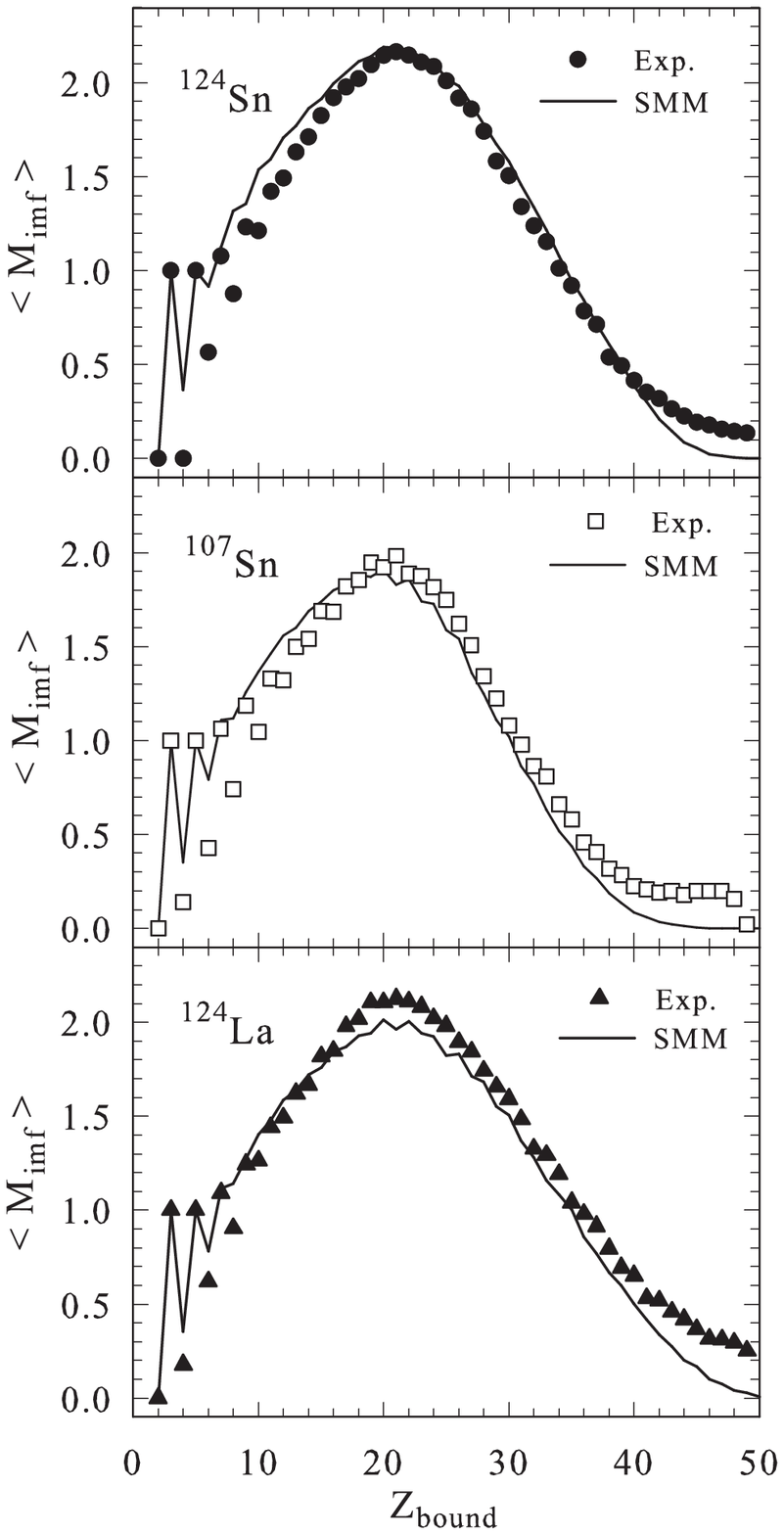}}
\caption{\small{Comparison of SMM calculations (lines) with the experimental
results (symbols) for the mean multiplicity $\langle M_{\rm imf} \rangle$ of intermediate-mass 
fragments versus $Z_{\rm bound}$ for the three studied projectiles.
}}
\label{fig:mimfsmm}
\end{figure} 

The results of the 
SMM calculations, represented by the thin lines in the 
figures, were obtained with the normalized ensembles (parameter $a_2 = 0.015$~MeV$^{-2}$, 
cf. Fig.~\ref{fig:eoa})
and with liquid-drop parameters that provide an adequate 
description of the experimental observables and trends. 
The charge observables are mostly sensitive to the surface-term coefficient $B_0$, as 
discussed in the previous section. In order to maintain a consistent comparison, the 
evidence for a reduced symmetry term coefficient $\gamma$ in the fragmentation channels, extracted
from the analysis of isotope characteristics in Sec.~\ref{sec:iso}, 
has already been taken into account
by choosing $\gamma = 14$~MeV for the three systems. 

For the surface-term coefficient, different values had to be chosen in order to obtain equivalent
descriptions for the neutron-rich and neutron-poor systems. The good reproduction of the charge
distributions (Figs.~\ref{fig:z107}--\ref{fig:zlan}) and of the charge correlations shown in 
Figs.~\ref{fig:zmaxsmm} and~\ref{fig:mimfsmm} has been achieved with the values $B_0 = 17.5$~MeV
for $^{124}$Sn and $B_0 = 19.0$~MeV for the neutron-poor $^{107}$Sn and $^{124}$La. They are
determined with an accuracy of about 0.5~MeV for the chosen ensemble of excited sources 
(cf. Figs.~\ref{fig:bonz} and \ref{fig:bonimf}). It should be mentioned here again that the 
comparisons are made with the measured data and that the small inefficiencies of the spectrometer 
for very light fragments are taken into account by filtering the calculations. If corresponding 
corrections are applied to the data, as done in Ref.~\cite{sfienti09}, the main effect is a
shift of the maxima of the mean fragment multiplicity to about 10\% larger values of 
$Z_{\rm bound}$ than shown in Fig.~\ref{fig:mimfsmm}, 
while the increase in the maximum multiplicity itself is nearly negligible.  

If modifications of the source parameters are simultaneously permitted,
the effect of $B_0$ on fragment multiplicities can be compensated by a corresponding variation
of $a_2$. The larger mean multiplicities $\langle M_{\rm imf} \rangle$ and the shift of the peak 
of the multiplicity distribution to larger $Z_{\rm bound}$ observed by reducing the surface term 
(Fig.~\ref{fig:bonimf}) can also be obtained by increasing the excitation energies of the
sources~\cite{Botvina06}. Equally satisfactory results for $^{124}$Sn were, e.g., obtained with the
parameter pair $a_2 = 0.015$~MeV$^{-2}$ and $B_0 = 17.5$~MeV used here as well as with
$a_2 = 0.012$~MeV$^{-2}$ and $B_0 = 19.0$~MeV. The required $N/Z$ dependence
of $B_0$ is independent of this ambiguity, at least as long as the ensemble parameters 
remain the same and independent of the projectile $N/Z$.

\begin{figure} [htb]
\centerline{\includegraphics[width=8cm]{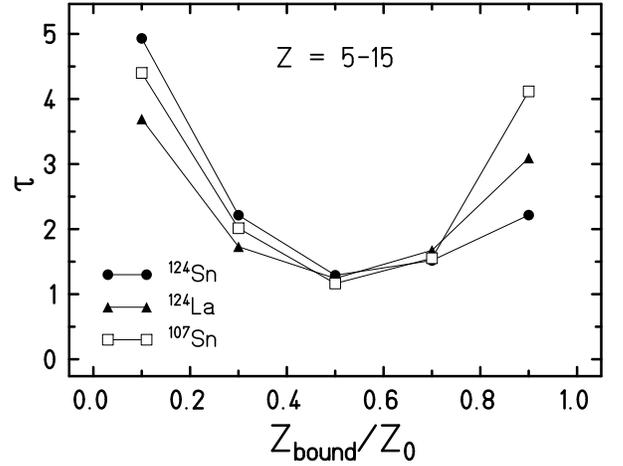}}
\caption{\small{Parameter $\tau$ obtained from power-law fits to the fragment yields $y(Z)$ 
in the range $5 \le Z \le 15$ according to $y(Z) \propto Z^{-\tau}$ 
as a function of $Z_{\rm bound}/Z_0$.
}}
\label{fig:tau}
\end{figure} 

Evidence for a variation of the surface-term coefficient with the isotopic composition of the 
system has previously been derived from a $\tau$ analysis of fragment-charge yields $y(Z)$ 
measured for the fragmentation of $^{129}$Xe, $^{197}$Au, and $^{238}$U at 600 
MeV/nucleon~\cite{Botvina06}. 
Near the onset of multifragmentation ($Z_{\rm bound}/Z_0 > 0.6$), the $\tau$ parameters obtained 
from power-law fits according to $y(Z) \propto Z^{-\tau}$ in the range $5 \le Z \le 15$ are 
larger for the projectiles with smaller $N/Z$ while the opposite is predicted by the SMM for 
constant $B_0$. There, the variation of the $N/Z$ of the studied systems was accompanied 
by a considerable change in mass. The 
present data confirm that the observed inversion of the hierarchy of $\tau$ with respect 
to the predictions with constant $B_0$ is indeed an isotopic effect. 
The $\tau$ parameters are larger for the two neutron-poor systems at large 
$Z_{\rm bound}/Z_0$ (Fig.~\ref{fig:tau}).
For the further conclusion that the $N/Z$ dependence of $B_0$ may disappear at higher 
excitations~\cite{Botvina06} no additional evidence is obtained from the present data.
In fact, the sensitivity of the $Z$ spectra to $B_0$ is seen to diminish at smaller 
$Z_{\rm bound}$ (Fig.~\ref{fig:bonz}) and the fragment multiplicities as a function 
of $Z_{\rm bound}$ become universal (Fig.~\ref{fig:bonimf}). 

A variation of the surface term with $N/Z$ is similar to a surface dependence 
of the symmetry energy, i.e. an additional term $\gamma_s \cdot A^{2/3}(A-2Z)^2/A^2$ in the 
liquid-drop parametrization. Using the effective $A/Z$ values of the two secondary beams
and $\Delta B_0 = 1.5$~MeV and averaging over the two pairs of systems,
$^{124}$Sn vs. $^{107}$Sn and $^{124}$Sn vs. $^{124}$La, yields a coefficient 
$\gamma_s = 45$~MeV. This is larger than parameters of standard mass 
formulas. That of Myers and Swiatecki, e.g., contains 28.1 and 33.2 MeV for the volume 
and surface coefficients of the symmetry term, respectively~\cite{ms}. 
The accuracy of the deduced $N/Z$ dependence of $B_0$ is not sufficient to draw firm 
conclusions at this point. However, large surface 
effects are consistent with the results for the symmetry term derived from the fragment 
mass yields discussed in the following section.

Concluding the analysis of charge observables, it may be stated that
an excellent overall description of the measured yields and correlations is obtained
within the statistical approach. With the chosen ensemble of excited spectator nuclei
and after a global normalization, the fragment yields and spectra shapes are reproduced on 
an absolute level (Figs.~\ref{fig:z107}-\ref{fig:zlan}) and the correlations agree 
well with the data (Figs.~\ref{fig:zmaxsmm} and~\ref{fig:mimfsmm}). Deviations occur for
very large $Z_{\rm bound}$ for which the yields of heavy fragments are overpredicted
and those of light fragments ($Z <$~10) are underpredicted.
They should not be overrated, however, because the incomplete trigger efficiency 
in the experiment (Sec.~\ref{sec:exp}) and the empirical character of the 
yield function of the theoretical source ensembles may both be causing them. In fact,
the finite values for the fragment multiplicity for $Z_{\rm bound}$ approaching 50 in the tin cases 
seem to indicate a trigger bias favoring the more violent collisions.

\section{Isotope characteristics of produced fragments}
\label{sec:iso}

\subsection{Parameter dependence}

\begin{figure} [tbh]
\centerline{\includegraphics[height=10cm]{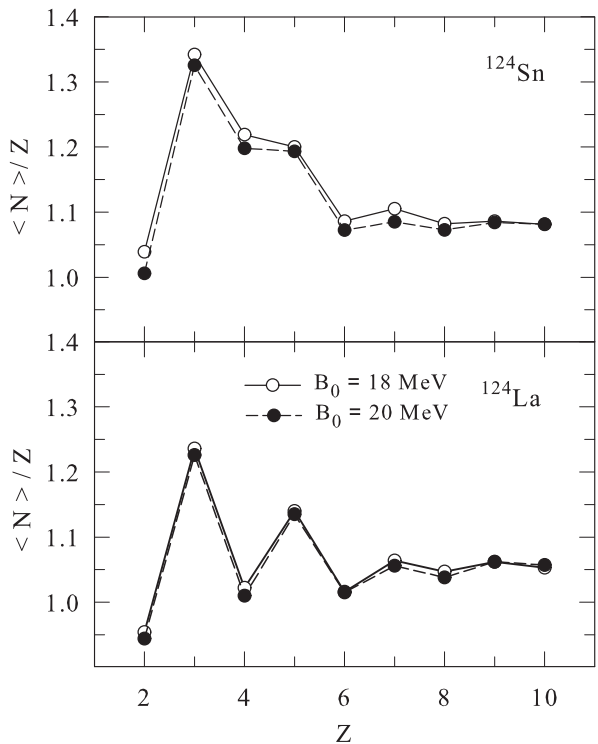}}
\caption{\small{SMM calculations of the mean neutron-to-proton ratio
$\langle N \rangle /Z$ of fragments with $2\le Z\le 10$ produced in the fragmentation
of $^{124}$Sn (top panel) and $^{124}$La (bottom panel) projectiles for the interval
$0.4 \le Z_{\rm bound}/Z_0 < 0.6$ as a function of the fragment $Z$, 
performed with $\gamma=25$~MeV 
and with two surface energy coefficients $B_0=$~18 and 20 MeV.
}}
\label{fig:nopsmm_B0}
\end{figure}

\begin{figure} [tbh]
\centerline{\includegraphics[height=10cm]{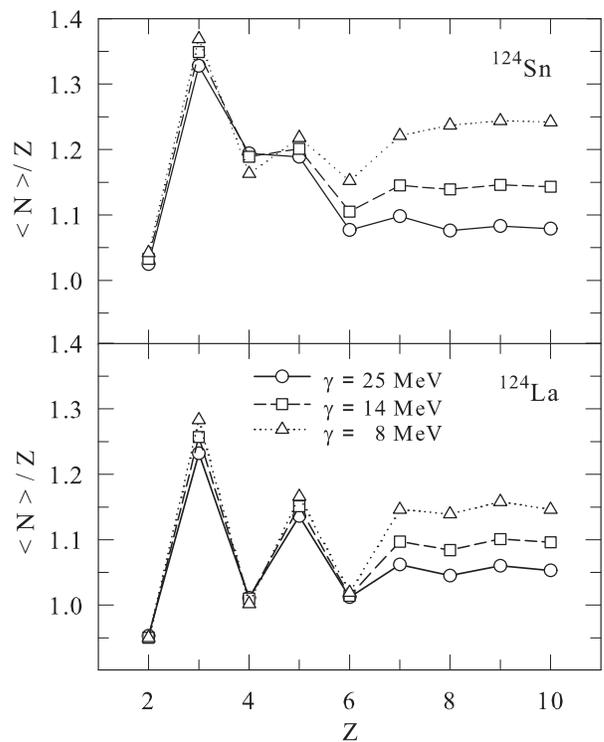}}
\caption{\small{SMM calculations of the mean neutron-to-proton ratio
$\langle N \rangle /Z$ as in Fig.~\ref{fig:nopsmm_B0}, but for $B_0=18$~MeV and 
three different symmetry energy coefficients $\gamma=$~8, 14, and 25 MeV.
}}
\label{fig:nopsmm_gamma}
\end{figure}

The successful description of charge observables provides the basis for the 
investigation of the isotope distributions and the information contained therein.
As in the previous section, the parameter dependence is studied first.
For this purpose, the mean neutron-to-proton ratios $\langle N \rangle /Z$ are chosen.
SMM calculations of this quantity for the two projectiles $^{124}$Sn and $^{124}$La 
and for different surface-term coefficients $B_0$ and different symmetry-term
coefficients $\gamma$ are presented in Figs.~\ref{fig:nopsmm_B0} 
and~\ref{fig:nopsmm_gamma}, respectively. 

As evident from the figures, the situation is complementary to that encountered 
in the study of the charge observables as noted previously~\cite{ogul09}.
The surface parameter $B_0$ has practically no influence on the isotope distributions
while the symmetry energy strongly affects the mean neutron richness of fragments, most
noticeably for $Z \ge 7$. It increases with decreasing
$\gamma$ because of the correspondingly larger widths of the isotope distributions at freeze-out.
At $Z \le 6$, the mean neutron-to-proton ratios exhibit the known odd-even effects, typical for 
this type of reaction~\cite{traut07nn} and shown to result
from secondary decays when the strongly bound even-even $N=Z$ nuclei attract a large fraction 
of the product yields~\cite{winch00,ricci04}.

The mean neutron numbers are larger for the fragments of $^{124}$Sn by about half a 
mass unit, and the difference is larger if $\gamma$ is reduced (Fig.~\ref{fig:nopsmm_gamma}).
This difference, however, is much smaller than that of the primary fragments, which
have nearly the same $\langle N \rangle /Z$ ratio as the primary thermal sources \cite{Botvina01}. 
During the secondary deexcitation, the neutron richness of fragments will decrease 
as neutron emission dominates. In the case of small $\gamma$, this trend is reduced
because the mass distribution of accessible daughter nuclei is wider and emissions
of charged particles may more easily compete~\cite{buyuk05}.

\subsection{Isotope distributions}

\begin{figure} [t]
\centerline{\includegraphics[width=8cm]{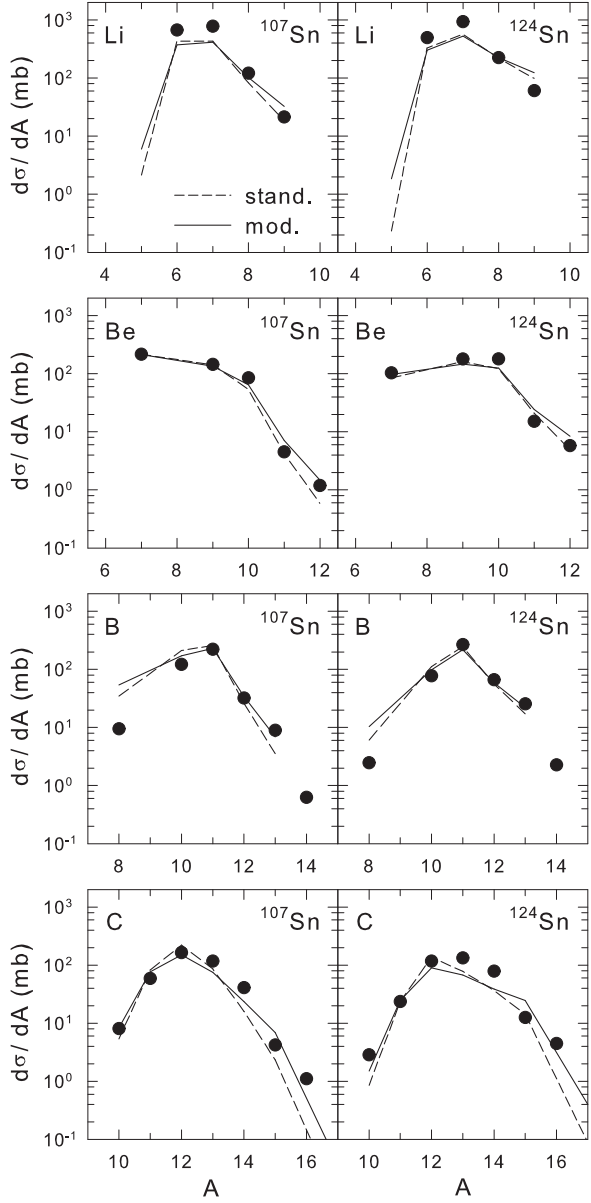}}
\caption{\small{Isotope distributions of fragments with $Z$~=~3--6 
for $^{107}$Sn (left panels) and $^{124}$Sn (right panels), integrated
over $0.2 \le Z_{\rm bound}/Z_0 \le 0.8$. 
The dots represent the experimental data. 
Normalized results from calculations with the standard and modified 
parameters are shown by the dashed and solid lines, respectively.
}}
\label{fig:adis_exp_smm_1}
\end{figure}

\begin{figure} [tbh]
\centerline{\includegraphics[width=8cm]{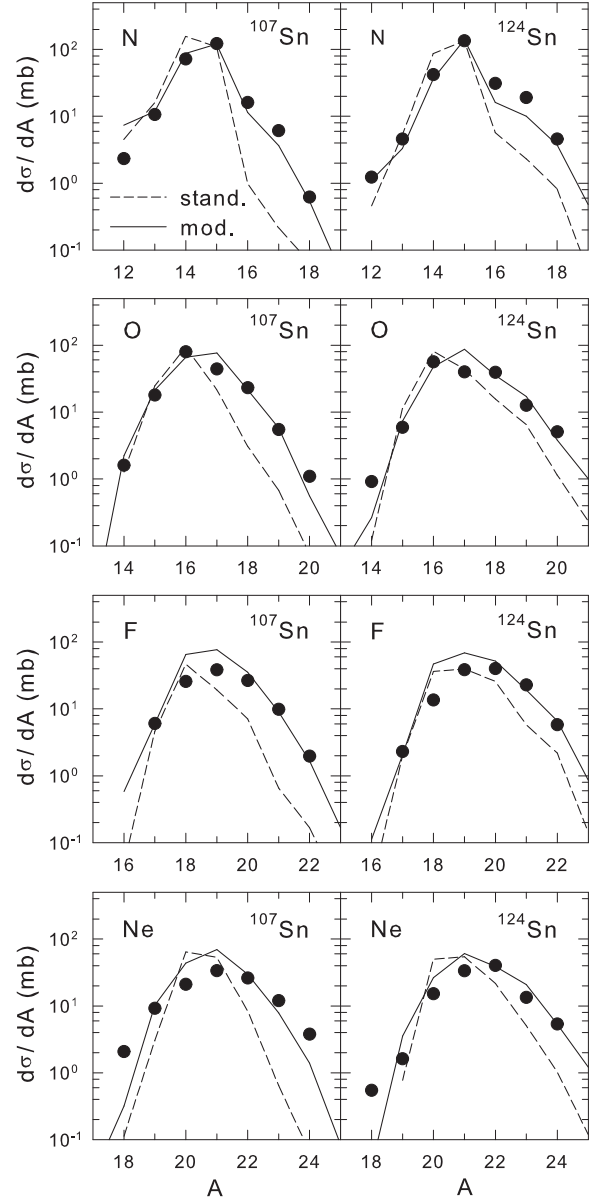}}
\caption{\small{Isotope distributions 
for $^{107}$Sn (left panels) and $^{124}$Sn (right panels)
as in Fig.~\protect\ref{fig:adis_exp_smm_1} but for
fragments with $Z$~=~7--10.
}}
\label{fig:adis_exp_smm_2}
\end{figure}

Cross sections measured for resolved isotopes and integrated over 
$0.2 \le Z_{\rm bound}/Z_0 \le 0.8$ are presented in Figs.~\ref{fig:adis_exp_smm_1} 
and \ref{fig:adis_exp_smm_2} for $Z$ = 3--6 and 7--10, respectively. 
The major part of the rise and fall of multifragmentation is covered by the selected 
$Z_{\rm bound}$ interval (Fig.~\ref{fig:mimfsmm}). 
Because of the stringent sorting conditions required for producing
background-free mass spectra, the experimental cross sections 
for isotope production were obtained by normalizing the mass-integrated fragment yields 
for a given $Z$ with respect to the measured elemental cross section in the same 
$Z_{\rm bound}$ interval (Sec.~\ref{sec:exp}). 
The differences of the isotope distributions between the neutron-poor
and neutron-rich cases are small but still visible in the logarithmic representation.
In particular, the fragment yields on the neutron-rich side are considerably larger
in the case of $^{124}$Sn.

The SMM calculations shown in the figures were performed with the ensemble 
parameter $a_2$ = 0.015 MeV$^{-2}$, with
$B_0$ = 17.5 and 19 MeV for the neutron-rich and neutron-poor projectiles, respectively,
and for two choices of $\gamma$. For the standard calculations, the symmetry-term coefficient
$\gamma = 25$~MeV was used for the hot fragments at breakup and throughout the secondary
deexcitation stage. Modified calculations were performed with $\gamma = 14$~MeV and with
the interpolation procedure for the secondary deexcitation at 
$E_x \le 1$~MeV/nucleon (Sec.~\ref{sec:sta}). 
The calculated cross sections are globally normalized for each projectile case as 
described in Sec.~\ref{sec:cha} A.

\begin{figure} [htb]
\centerline{\includegraphics[width=8.0cm]{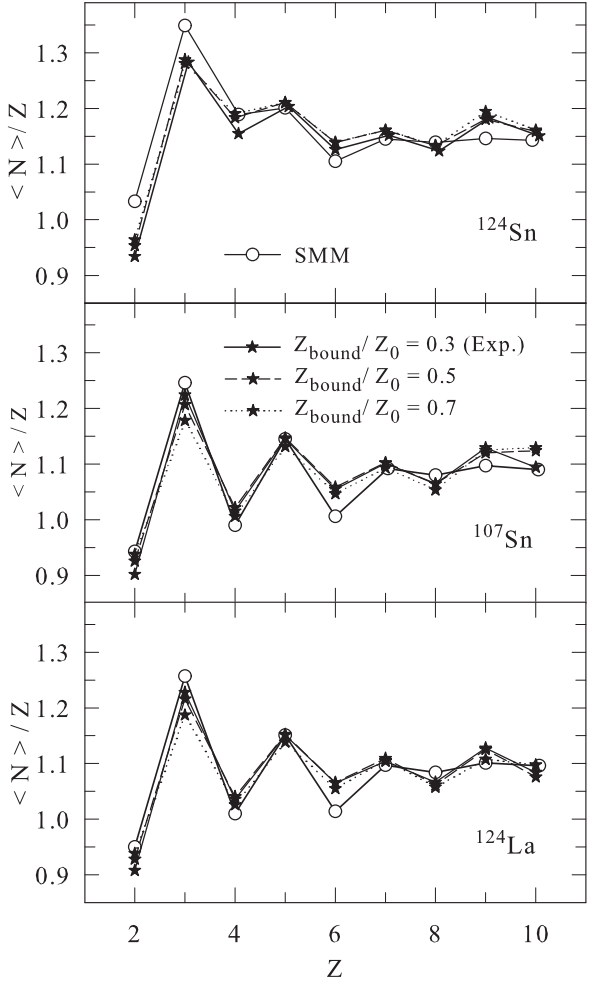}}
\caption{\small{Measured mean neutron-to-proton ratios $\langle N \rangle /Z$ for fragments with
$Z \le 10$ produced in the multifragmentation of $^{124}$Sn (top panel), $^{107}$Sn (middle panel), 
and $^{124}$La projectiles (bottom panel) and for three intervals of $Z_{\rm bound}/Z_0$ with centers 
as indicated and widths 0.2 (stars).
In the calculations, performed for the range $0.2 \le Z_{\rm bound}/Z_0 < 0.8$,
the surface-energy coefficients $B_0$=18 MeV and the 
symmetry-energy coefficient $\gamma =14$~MeV were used (open circles).
}}
\label{fig:noverz}
\end{figure} 

The comparison with the data shows that the yields of neutron-rich isotopes are much better
described if the reduced symmetry-energy coefficient is used. In particular, the heavier
fragments with $Z \ge$~7 are rather sensitive to the variation of $\gamma$, as already evident from
Fig.~\ref{fig:nopsmm_gamma}. 
The same conclusion can be drawn from the comparison of the mean neutron-to-proton ratios 
$\langle N\rangle /Z$ for elements of $Z \le 10$ with the experimental results (Fig.~\ref{fig:noverz}).
Evidently, they depend very little on $Z_{\rm bound}$, and the calculations are,
therefore, only shown for the full range $0.2 \le Z_{\rm bound}/Z_0 < 0.8$.
The modified calculations, with $\gamma=14$~MeV, agree well with the data, both with regard to the 
average magnitude as well as to the odd-even structures as a function of $Z$. The latter,
however, are slightly more pronounced 
in the calculations than in the experiment for $Z \le 6$ while the
opposite is the case for larger fragments. This small difference may be connected to the
discontinuity caused by the mass limit $A \le 16$ set for using the Fermi breakup in the 
secondary-decay part of the calculations.

\begin{figure} [tbh]
\centerline{\includegraphics[width=8.0cm]{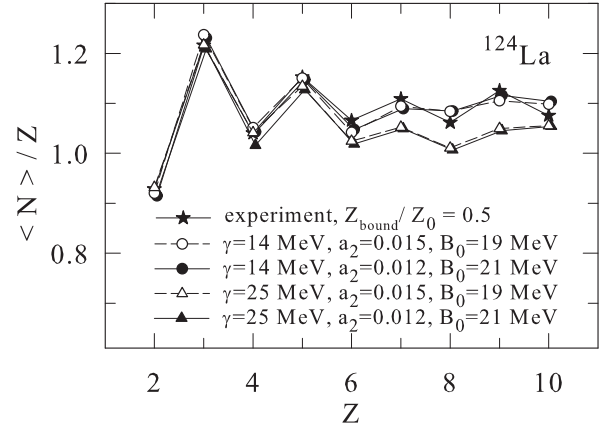}}
\caption{\small{Mean neutron-to-proton ratio $\langle N \rangle /Z$ of fragments with $Z \le 10$
produced in the fragmentation of $^{124}$La projectiles 
in the range $0.4 \le Z_{\rm bound}/Z_0 < 0.6$.
The experimental results (stars) are compared with SMM calculations using four combinations of 
symmetry-energy, ensemble, and surface parameters as indicated. Only the calculations with the
symmetry-energy coefficient $\gamma = 14$~MeV (circles) are close to the data.
}}
\label{fig:a12}
\end{figure}

The ambiguity in the choice of the ensemble and surface-term parameters $a_2$ and $B_0$ discussed in 
Sec.~\ref{sec:cha}.C does not affect this result. In Fig.~\ref{fig:a12}, calculations are shown
that were performed with several combinations of these parameters, all capable of reproducing 
the charge observables. The observed mean $\langle N \rangle /Z$ is only reached 
with the reduced symmetry-term coefficient $\gamma = 14$~MeV. 

\begin{figure} [tbh]
\centerline{\includegraphics[width=8.0cm]{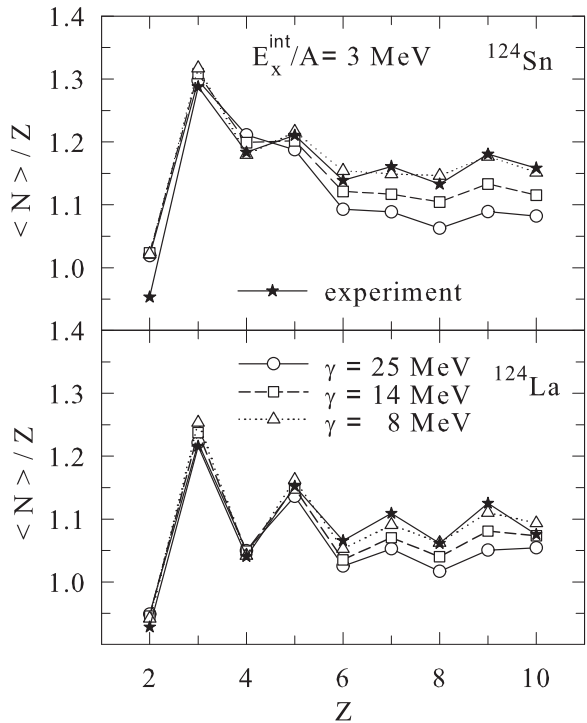}}
\caption{\small{Mean neutron-to-proton ratio $\langle N \rangle /Z$ of fragments
produced in the fragmentation of $^{124}$Sn (top panel) and $^{124}$La (bottom panel) 
projectiles in the range $0.4 \le Z_{\rm bound}/Z_0 < 0.6$. 
The experimental results (stars) are compared with SMM calculations using 
the three indicated values of the symmetry-term coefficient $\gamma$ and 
$E_x^{\rm int} = 3$~MeV/nucleon as the start value of the interpolation interval for the 
secondary-decay stage of the calculations. With this choice, $\gamma = 8$~MeV (triangles) 
gives the best agreement with the data.
}}
\label{fig:ex_int}
\end{figure}

Another important parameter is related to the secondary-deexcitation stage of the calculations.
For the restoration of the standard symmetry-term coefficient, a linear interpolation between 
the initially reduced value and $\gamma = 25$~MeV of isolated nuclei in the interval of excitation 
energies below $E_x^{\rm int} = 1$~MeV/nucleon has been adopted (end of Sec.~\ref{sec:sta}). 
The evaporation times corresponding to this energy are of the order of 10$^{-21}$~s which seems long,
and one may argue that the restoration should start at a higher initial value $E_x^{\rm int}$, 
possibly closer to 3~MeV/nucleon which would correspond to a thermal freeze-out 
at $T \approx 5$~MeV~\cite{traut08}. 
Ensemble calculations performed with this choice for the $^{124}$Sn and $^{124}$La systems also
provide an excellent description of the data (Fig.~\ref{fig:ex_int}). It is even superior in the 
range $Z \ge 6$ where the odd-even staggering of $\langle N \rangle /Z$ is much better reproduced 
than with $E_x^{\rm int} = 1$~MeV/nucleon (cf. Fig.~\ref{fig:noverz}).
It requires, however, an even larger reduction of the symmetry term coefficient $\gamma$ to values
below 10 MeV. Therefore, to the extent that $E_x^{\rm int} = 1$~MeV/nucleon has to be considered 
as a lower limit for the beginning restoration of the properties of isolated nuclei, 
the reduced $\gamma = 14$~MeV 
represents an upper limit for the symmetry-term coefficient needed to reproduce the 
observed neutron richness of intermediate-mass fragments. 

Summarizing this section it may be stated that the isotopically resolved fragment
yields provide evidence for a reduced symmetry energy in the hot environment at
freeze-out, in accordance with previous findings~\cite{LeFevre,Botvina05}.
Similar observations were made in reaction studies performed at intermediate 
and relativistic energies~\cite{iglio06,souliotis07,ogul09,hudan09,henzlova10}.

\begin{figure} [tbh]
\centerline{\includegraphics[width=8.0cm]{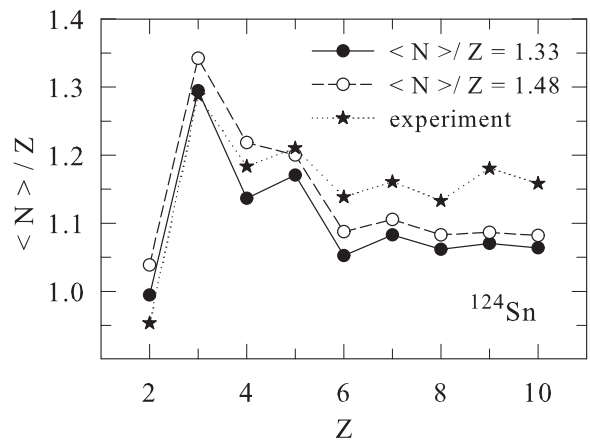}}
\caption{\small{Mean neutron-to-proton ratio $\langle N \rangle /Z$ of fragments with $Z \le 10$
produced in the fragmentation of $^{124}$Sn projectiles 
in the range $0.4 \le Z_{\rm bound}/Z_0 < 0.6$. 
The experimental results (stars) are compared with SMM calculations for the 
nominal $N/Z =$~1.48 of that system (open circles) and for a reduced initial $N/Z =$~1.33 
ratio (dots). The standard liquid-drop parameters $B_0$ = 18 and $\gamma$ = 25 MeV were used.
}}
\label{fig:noverz_2}
\end{figure} 

\section{Discussion of freeze-out properties}
\label{sec:dis}

\subsection{Source properties}

Several reasonable but also essential assumptions have been made in the statistical 
analysis presented so far. In particular, for the $N/Z$ ratios of the considered ensembles
those of the initial projectiles have been used. At relativistic energies, the preequilibrium 
particles are mainly produced in individual nucleon-nucleon collisions and, on average, the 
neutron-to-proton ratios should not change. 
This is contrary to the situation in the intermediate-energy range (beam energies less than 
100 MeV/nucleon) where a considerable isospin diffusion has been identified \cite{tsang04}.
Studying this in more detail with INC or BUU 
calculations shows that, at relativistic energies, a slight decrease of the neutron richness 
with increasing excitation energy has to be expected. It is not strongly dependent on the 
initial $N/Z$ so that the 
difference between the isotopic compositions of different residue systems does not change 
by more than a few percent during the initial reaction stages~\cite{Botvina02,LeFevre}. 
This result is essential for the conclusions drawn from the isoscaling study of 
Le F\`{e}vre {\it et al.}~\cite{LeFevre} as well as for the isoscaling 
analysis presented further below. 
The isoscaling parameter $\alpha$ is roughly proportional to the difference of the $N/Z$ ratios of
the studied systems so that a reduction of the latter, before the chemical fragment composition is 
established, would naturally explain the observed decrease of $\alpha$.

The situation is different for the analysis of the $\langle N \rangle /Z$ ratios of light 
fragments presented in the previous section, as illustrated in Fig.~\ref{fig:noverz_2}. 
Two calculations were performed for $^{124}$Sn ensembles, one with the
original $N/Z = 1.48$ of this projectile and one with a considerably smaller neutron-to-proton ratio 
$N/Z = 1.33$ which is halfway between those of the $^{124}$Sn and of the neutron-poor $^{124}$La and 
$^{107}$Sn nuclei. As expected, the $\langle N \rangle /Z$ ratios of the produced fragments are 
lower for the smaller initial $N/Z$ than for the original choice. To raise them to the 
experimental level an even larger reduction of the $\gamma$ parameter would be necessary. 
A larger than assumed variation of the isotopic composition of the thermal sources at 
breakup will thus not affect the previously reached conclusion.

\subsection{Freeze-out temperature}

\begin{figure}[t]

   \centerline{\includegraphics[width=8.0cm]{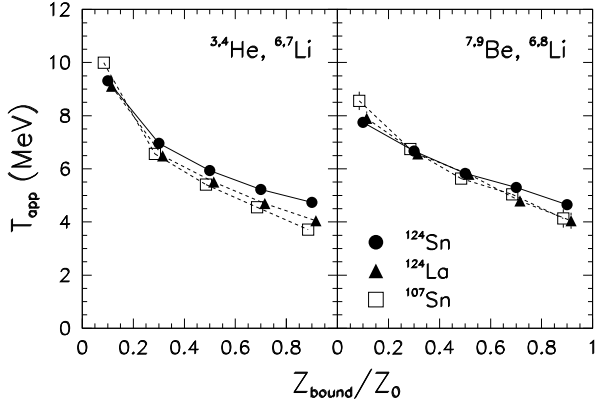}}

\caption{\small{Apparent temperatures $T_{\rm HeLi}$ (left panel) and $T_{\rm BeLi}$ 
(right panel) as a function of $Z_{\rm bound}/Z_0$ for the three reaction systems.
For clarity, two of the three data sets are slightly
shifted horizontally; statistical errors are displayed where they are larger than
the symbol size (from Ref.~\protect\cite{sfienti09}).
}}
\label{fig:temp_exp}
\end{figure}

Chemical freeze-out temperatures, as obtained from double-isotope-yield ratios measured in this
experiment, have been presented and discussed in Ref.~\cite{sfienti09}. 
The two observables shown in Fig.~\ref{fig:temp_exp} as a function of the normalized
$Z_{\rm bound}$ are the frequently used
\begin{equation}
T_{\rm HeLi} = 13.3~{\rm MeV}/\ln(2.2\frac{Y_{^{6}{\rm Li}}/Y_{^{7}{\rm Li}}}
{Y_{^{3}{\rm He}}/Y_{^{4}{\rm He}}}),
\label{eq:heli}
\end{equation}
based on the $^3$He/$^4$He and $^6$Li/$^7$Li yield ratios where $Y$ denotes the isotopically 
resolved yields (left panel, Ref.~\cite{poch95}), and 
\begin{equation}
T_{\rm BeLi} = 11.3~{\rm MeV}/\ln(1.8\frac{Y_{^{9}{\rm Be}}/Y_{^{8}{\rm Li}}}
{Y_{^{7}{\rm Be}}/Y_{^{6}{\rm Li}}}),
\label{eq:beli}
\end{equation}
the latter being deduced from Li and Be fragment yields (right panel, Ref.~\cite{traut07}). 
The apparent temperatures as given by the formulas are displayed,
i.e. without corrections for secondary decays feeding the ground states of these nuclei.
Secondary decays are expected to lower the apparent values by 10--20\% with respect to the 
actual breakup temperature~\cite{poch95,traut07}. 

In Ref.~\cite{sfienti09}, the near invariance of the measured breakup temperatures with the 
isotopic composition of the fragmenting system was shown to be contrary to the Hartree-Fock 
predictions for the limiting temperatures of excited compound nuclei~\cite{besp89}. 
Their global behavior is, on the other hand, in good qualitative  
agreement with the SMM calculations for $^{124}$Sn and $^{124}$La nuclei of Ogul and 
Botvina~\cite{ogul02}. 
The differences obtained for these two cases are negligible in the 
multifragmentation regime and reach a maximum $\Delta T \approx 0.4$~MeV in the transition 
region where the equilibrium temperature for the more proton-rich $^{124}$La system
is slightly lower. As discussed in Ref.~\cite{ogul02}, 
the difference appears because, in the case of neutron-rich sources, 
partitions with a compound-like fragment and many free neutrons are still frequent in the 
transition regime. Their temperature is higher than that of partitions involving a few charged
fragments which are more probable for neutron-poor systems. 

\begin{figure} [t]
\centerline{\includegraphics[width=8cm]{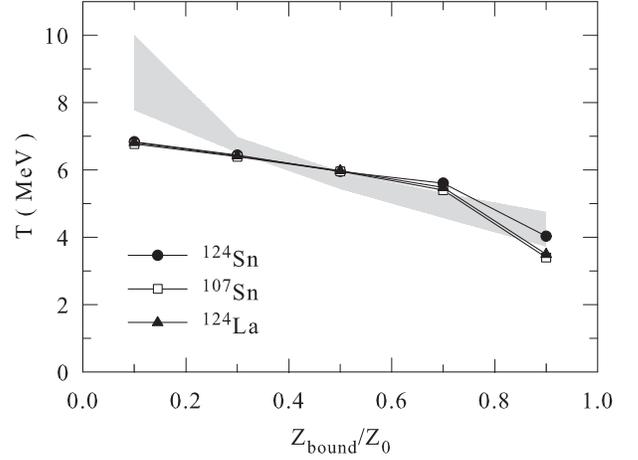}}
\caption{\small{Mean microcanonical temperatures calculated for the three projectile ensembles 
with the $N/Z$ dependent surface coefficients $B_0 =$~17.5 and 19~MeV and with a symmetry 
coefficient $\gamma = 14$~MeV.
The experimental result is represented by the shaded band that extends from the lowest
to the highest of the six apparent $T_{\rm HeLi}$ and $T_{\rm BeLi}$ temperatures shown in
Fig.~\protect\ref{fig:temp_exp}. 
}}
\label{fig:temp_smm}
\end{figure} 

The ensemble calculations performed in the present work permit a more realistic comparison than 
those for sources with fixed mass and charge of Ref.~\cite{ogul02}. 
The result, however, remains the same. 
As shown in Fig.~\ref{fig:temp_smm}, the dependence on $N/Z$ of the mean microcanonical breakup 
temperatures, sorted according to $Z_{\rm bound}$, is negligible in the regime of 
multifragmentation. In the most peripheral bin with large $Z_{\rm bound}$, the temperature of 
the more neutron-rich system is slightly larger, a tendency that is also apparent in the 
experimental isotope temperatures (Fig.~\ref{fig:temp_exp}).
The shaded band in the figure represents the full range of the apparent $T_{\rm HeLi}$ and 
$T_{\rm BeLi}$ temperatures which, in the regime of multifragmentation, are close to the
microscopic temperatures. These results not only confirm the usefulness of isotope 
temperatures for describing properties of the 
chemical freeze-out stage but also suggest that the transition from compound decay to 
multifragmentation can be explained statistically with the opening of the fragmentation 
phase space~\cite{sfienti09}. 

The calculations were performed with the surface and symmetry-energy parameters suggested by the
analyses of the previous sections, in particular also with $\gamma = 14$~MeV. 
Variations of this parameter did not influence the results, however~\cite{buyuk05}.
On the absolute scale, the microscopic temperature of 6.0 MeV in the center of the fragmentation 
regime ($\langle Z_{\rm bound}/Z_0 \rangle$ = 0.5) is only a few percent higher than the 
apparent isotope temperatures. 
The present calculations thus suggest less of a side-feeding correction than 
the up to $\approx$~20\% derived from a variety of models in Refs.~\cite{poch95,traut07}. 
This emphasizes once more the quantitative uncertainty inherent in estimating the effects of 
secondary decays but also supports the conclusion that they are relatively small for the two 
temperature observables chosen here. 
The calculations do not reproduce the accelerated rise of the apparent 
temperature $T_{\rm HeLi}$ at very small $Z_{\rm bound}/Z_0 \le 0.2$ and are still about 1 MeV 
below the corresponding value of $T_{\rm BeLi}$. A contribution to the event class with very small 
$Z_{\rm bound}$ from collisions with more central impact parameters can explain this observation
(cf. Sec.~\ref{sec:exp}). Isotope 
temperatures involving yields of carbon and oxygen isotopes have also been investigated and 
found to remain close to $T \approx 6$~MeV, practically independent of $Z_{\rm bound}$ and in 
good agreement with the results reported for the fragmentation of $^{197}$Au
projectiles~\cite{traut07}.

\subsection{Isoscaling and the symmetry term}

\begin{figure} [hbt]
\centerline{\includegraphics[height=11.5cm]{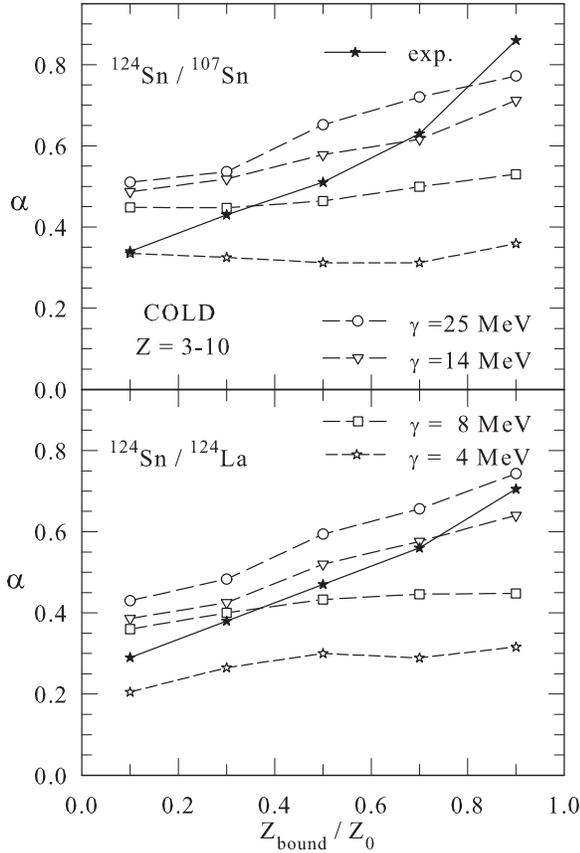}}
\caption{\small{Experimental data (stars) and SMM ensemble calculations (open symbols) of
isoscaling coefficients $\alpha$ extracted from fragment yield ratios ($3 \le Z \le 10$) 
for $^{124}$Sn and $^{107}$Sn projectiles (top panel) and for $^{124}$Sn and $^{124}$La
projectiles (bottom panel) as a function of the reduced bound charge $Z_{\rm bound}/Z_0$. 
Four different symmetry-term coefficients $\gamma$ were used 
in the SMM calculations as indicated.
}}
\label{fig:alpha_exp}
\end{figure} 

Isoscaling concerns the production ratios $R_{21}$ of fragments with
neutron number $N$ and proton number $Z$ in reactions with different
isospin asymmetry. It is defined as their exponential dependence on
$N$ and $Z$ according to
\begin{equation}
R_{21}(N,Z) = Y_2(N,Z)/Y_1(N,Z) \propto exp(N\cdot \alpha + Z\cdot
\beta)
\label{eq:scal}
\end{equation}
with parameters $\alpha$ and $\beta$. Here $Y_2$ and $Y_1$ denote
the yields from the more neutron-rich and the more neutron-poor reaction
system, respectively~\cite{Botvina02,tsang01}. Isoscaling has the property that a comparison 
is made for the same isotopes produced by different sources, so that the effects of yield 
fluctuations due to nuclear structure effects may be reduced.

\begin{figure} [bth]
\centerline{\includegraphics[height=11.5cm]{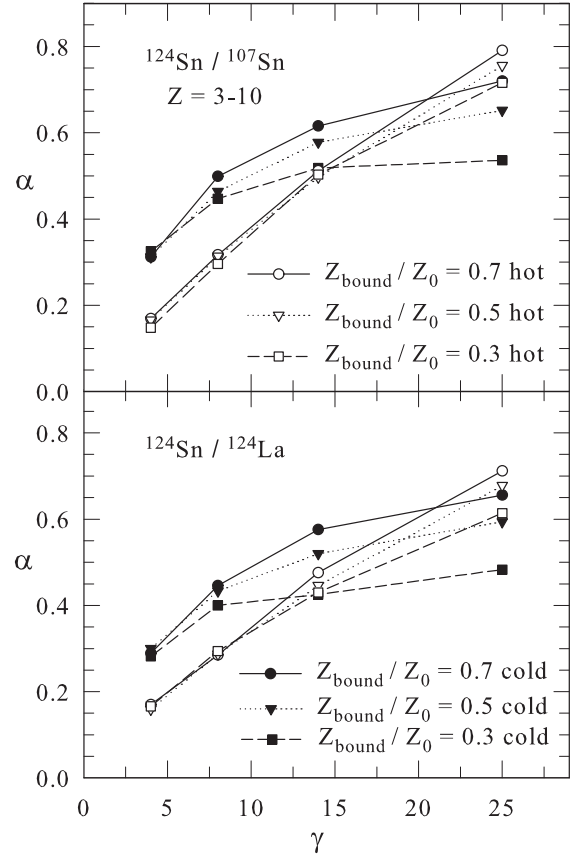}}
\caption{\small{Theoretical isoscaling coefficients $\alpha$ calculated for three 
intervals in $Z_{\rm bound}/Z_0$ and for
both hot primary fragments (open symbols, legend in upper panel) and cold fragments
after secondary deexcitation (closed symbols, legend in lower panel). 
}}
\label{fig:alpha_smm}
\end{figure} 

The comparison of the measured isotope yields from the fragmentation of $^{124}$Sn and of the
two neutron-poor systems has confirmed that isoscaling is observed~\cite{traut08,bianchin_bormio}. 
The isoscaling parameter $\alpha$, determined from the yields for $3 \le Z \le 10$ 
in different ranges of $Z_{\rm bound}/Z_0$ is shown in Fig.~\ref{fig:alpha_exp}. 
It is seen to decrease rapidly as the disintegration of the spectator 
systems into fragments and light particles increases, 
as reported previously for the fragmentation of target spectators in reactions of 
$^{12}$C on $^{112,124}$Sn at 300 and 600 MeV/nucleon~\cite{LeFevre}. 
Nearly identical results are obtained for the isotopic and isobaric pairs of reactions.

In the same figure, the results obtained from SMM ensemble calculations with different 
symmetry-term coefficients $\gamma$ and determined for $3 \le Z \le 10$ are compared with 
the data. Smaller values of $\gamma$ lead to
smaller isoscaling parameters $\alpha$ because the isotope distributions become wider and the 
variation of the yield ratios with $N$ becomes correspondingly smaller. The SMM standard value 
$\gamma$ = 25 MeV is applicable only in the bin of largest $Z_{\rm bound}$. Smaller values must 
be chosen to reproduce the rapidly decreasing parameter $\alpha$ in the fragmentation 
regime at smaller $Z_{\rm bound}$.

The proportionality expected for the dependence of $\alpha$ on $\gamma$ is demonstrated in 
Fig.~\ref{fig:alpha_smm}. It is fulfilled in good approximation for the hot fragments at freeze-out
(open symbols in the figure). 
Using the calculated values of $\alpha$ for $\gamma = 14$~MeV and the microscopic temperature
for this interval, $T = 6.0$~MeV, it is even possible to test the widely used formula
\begin{equation} 
\alpha \approx \frac{4\gamma}{T}
(\frac{Z_{1}^2}{A_{1}^2}-\frac{Z_{2}^2}{A_{2}^2}),
\label{eq:alph}
\end{equation}
where $Z_{1}$,$A_{1}$ and $Z_{2}$,$A_{2}$ are the atomic and mass
numbers of the two thermalized systems. The numerical factor deduced from the
ensemble calculations is 3.8 while the nominal coefficient, analytically derived in the
zero-temperature limit, is 4~\cite{Botvina02}. 
The validity of Eq.~\ref{eq:alph} at finite temperatures has
recently also been confirmed by Souza {\it et al.} in a theoretical study with the
SMM~\cite{souza09}.

\begin{figure} [tbh]
\centerline{\includegraphics[width=8cm]{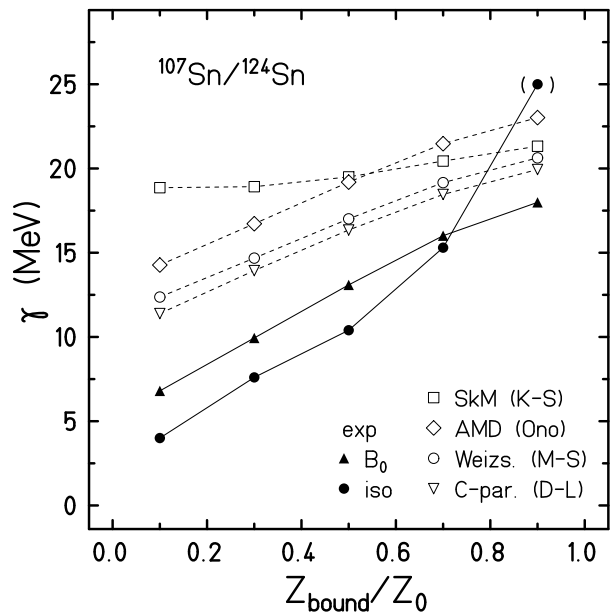}}
\caption{\small{Effective symmetry term coefficient $\gamma$ as expected from the changing
fragment-mass distributions using surface and volume symmetry-term coefficients
from Refs.~\protect\cite{ms,ono04,dani09,kol08} (open diamonds, circles, triangles, 
and squares, respectively) in comparison with 
the coefficient $\gamma$ for hot fragments obtained with the SMM 
from the $N/Z$ dependence of $B_0$ (solid triangles) and from the isoscaling
analysis for the $^{107,124}$Sn pair of reactions (dots).
}}
\label{fig:ghot}
\end{figure} 

This relation was independently deduced from dynamical and statistical 
investigations~\cite{Botvina02,tsang01a,ono03}
and used in several analyses of experiments \cite{LeFevre,iglio06,souliotis07,kowalski07}.
The SMM ensemble calculations show that, within some approximation, this relation remains valid for the
cold fragment yields as long as $\gamma$ is close to its conventional value of 25 MeV. 
As $\gamma$ is lowered
the resulting $\alpha$ for the asymptotic fragments becomes larger and deviates increasingly from the
value for hot fragments (Fig.~\ref{fig:alpha_smm}). During the late stages of the secondary decays,
the available daughter states are increasingly concentrated in the valley of stability which narrows
the initially wide isotope distributions. This effect has been demonstrated already and discussed in 
Ref.~\cite{LeFevre}.

\subsection{The role of the surface}

The symmetry-term coefficient $\gamma$ resulting from the isoscaling analysis is
shown in Fig.~\ref{fig:ghot} (dots) as a function of the sorting variable $Z_{\rm bound}$.
For large $Z_{\rm bound}/Z_0 >0.8$, the experimental values of $\alpha$ for the isotopic 
and isobaric pairs of reactions differ (Fig.~\ref{fig:alpha_exp}) but the prediction
for $\gamma = 25$~MeV is close to the mean. The rapid drop for smaller 
$Z_{\rm bound}$ is consistently evident from the two data sets (Fig.~\ref{fig:alpha_exp}).
For comparison, four predictions are shown in Fig.~\ref{fig:ghot}, obtained with rather different 
approaches but
all containing the effect of the surface-symmetry term whose importance increases for the 
lower-mass fragments~\cite{souza08}. 
The coefficients of the mass formula of Myers and Swiatecki~\cite{ms} are
adapted to ground-state masses, and values close to them have been used in
other studies~\cite{rad_gulm07}. From the energies of isobaric analog 
states, a relation between the volume and surface capacitances of nuclei for absorbing
asymmetry $N-Z$ was derived by Danielewicz and Lee~\cite{dani09}, while Kolomietz and Sanzhur
have used a variational approach using Skyrme forces to derive equilibrium values for the
volume symmetry term with surface and curvature corrections for nuclei along the 
$\beta$-stability line~\cite{kol08}. Theoretical surface and volume terms were also
obtained by Ono {\it et al.}~\cite{ono04} from fitting the ground-state binding energies 
calculated with the antisymmetrized molecular dynamics (AMD) model. 

With these parametrizations, an effective symmetry energy 
averaged over the set of partitions was calculated for the five bins in $Z_{\rm bound}$
after the experimental $Z$ distributions had been converted to mass distributions using 
the projectile $N/Z$. The obtained results show similar trends. 
The smaller fragments produced at higher excitations cause the effective mean 
symmetry term to decrease with decreasing $Z_{\rm bound}$ in all four cases 
but at a slower rate than that resulting from the isoscaling analysis of the 
experimental yield ratios.

The same procedure was, in addition, also performed with the surface-symmetry coefficient
$\gamma_s = 45$~MeV derived from the $N/Z$ dependence of the surface-term coefficient $B_0$
(Sec.~\ref{sec:cha}), combined with the volume-symmetry coefficient $\gamma_v = 28$~MeV of 
the mass formula of Myers and Swiatecki~\cite{ms}. 
As expected, the drop with decreasing $Z_{\rm bound}$ is faster
because of the larger surface-symmetry coefficient and close to that deduced from the 
isoscaling analysis. Within their larger uncertainty, the conclusions drawn from the analysis 
of the charge observables are thus consistent with those obtained from the isoscaling analysis
of the measured isotope distributions. Increased surface effects are expected if deformations and
exotic shapes are present among the smaller fragments at the chemical freeze-out. 

The symmetry-term coefficients of the AMD ground-state nuclei follow the
trend exhibited by the other mass formulas. However, for excited nuclei at the 
breakup stage, a reduced volume term and an isospin-independent surface term 
for the larger IMF's were found in the AMD study~\cite{ono04}. 
From this result and from the $\tau$ analysis of fragment charge yields~\cite{Botvina06} 
one may conclude
that the strong decrease of $\gamma$ with decreasing $Z_{\rm bound}$ is unlikely
to result from a surface effect alone. In fact, the surface term should decrease also 
for fragments in close vicinity of other nuclei because of residual interactions. The reduced 
symmetry term will then have to originate from a decrease of the overall matter
density caused by the expansion of the system~\cite{li_chen_06,buyuk08} 
and leading to expanded or deformed shapes of fragments as they are produced. 
These effects should be considered in the AMD model but were, perhaps,
not tested over the full range of possible excitations in Ref.~\cite{ono04}. There, the 
temperature at breakup was estimated to be less than 4 MeV which corresponds to the
bin of largest $Z_{\rm bound}$ here (Fig.~\ref{fig:temp_exp}) for which
the deduced $\gamma$ is comparable to the AMD result. The stronger reduction is
observed at the higher temperatures at smaller $Z_{\rm bound}$ (Fig.~\ref{fig:ghot}).

On the quantitative level, different degrees of reduction are deduced with the different methods.
The lowest values for $\gamma$ are obtained from the isoscaling analysis. 
The $\gamma \approx 10$~MeV obtained for the center of the fragmentation regime, 
i.e. $Z_{\rm bound}/Z_0 \approx 0.5$, is to be compared to 13 MeV, 
corresponding to $\Delta B_0 = 1.5$~MeV but carrying a large uncertainty, 
and to $\gamma = 14$~MeV which provided the best
global agreement with the charge and isotope observables studied in Secs.~\ref{sec:cha} 
and~\ref{sec:iso}. An answer may come from the fact that isoscaling is more sensitive 
to the widths of the isotope distributions and thus may give a larger weight to their tails.
These quantities, on the other hand, may more strongly depend on the realization chosen for the
secondary decay stage of the reaction. As discussed above, values below 14~MeV 
are favored by the odd-even structure of $\langle N \rangle /Z$ for $Z \ge 6$ (Fig.~\ref{fig:ex_int}).

\begin{figure} [tbh]
\centerline{\includegraphics[width=8cm]{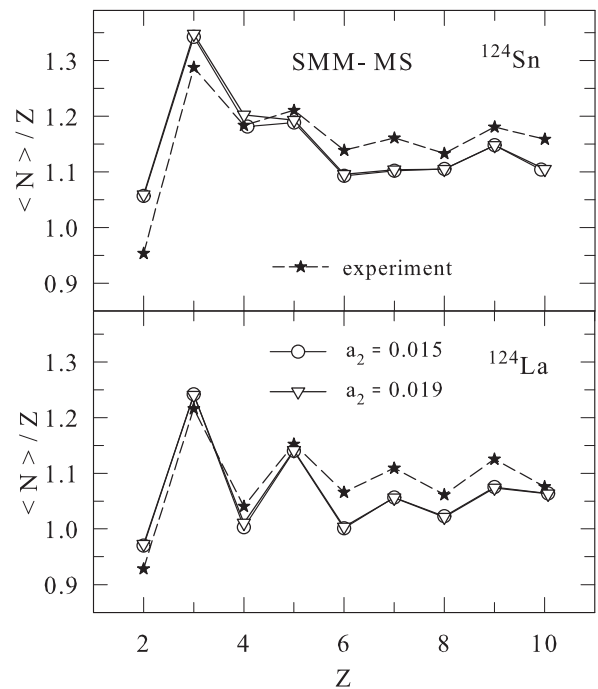}}
\caption{\small{Mean neutron-to-proton ratio $\langle N \rangle /Z$ of fragments with $Z \le 10$
produced in the fragmentation of $^{124}$Sn (top panel) and $^{124}$La (bottom panel) 
projectiles in the range $0.4 \le Z_{\rm bound}/Z_0 < 0.6$.
The experimental results (stars) are compared with SMM calculations 
with the Myers-Swiatecki parametrization~\protect\cite{ms} for the 
liquid-drop description of fragments and for two ensembles as indicated (open symbols). 
}}
\label{fig:noverz_ms}
\end{figure} 

The global $\gamma = 14$~MeV seems close to the effective coefficient $\gamma = 17$~MeV 
obtained from the parametrizations of Myers and Swiatecki or Danielewicz and Lee, 
including only surface effects of isolated nuclei (Fig.~\ref{fig:ghot}). However, the difference
is significant.
SMM calculations using the coefficients of Myers and Swiatecki
for the fragment description throughout, i.e., at breakup as well as during the secondary decay 
stages, lead to significantly smaller $\langle N \rangle /Z$, as shown for the $^{124}$Sn and
$^{124}$La systems in Fig.~\ref{fig:noverz_ms}. They are not 
sufficient to reproduce the observed neutron richness of light fragments. When searching for 
possible reasons for these large differences it was found that the reduction of $\gamma$
lowers the $Q$-value for proton emission while this is not the case with the Myers-Swiatecki
parameters. The relative probability for proton emission is thus enhanced, leading to more
neutron-rich daughter nuclei.

\section{Conclusions}
\label{sec:con}

New results from a systematic study of the $N/Z$ dependence of projectile fragmentation at
an incident energy of 600 MeV/nucleon have been presented. By employing radioactive secondary
beams, a large range of isotopic compositions has been explored, up to the present technical
limit. Global fragmentation observables were found to depend weakly on the projectile $N/Z$ while
substantial differences are observed for the isotope distributions of light fragments ($Z \le 10$).

The significance of the measured charge and mass yields and correlations was investigated with 
the SMM by performing ensemble calculations accounting for the
mass-energy correlation of the produced spectator nuclei. The ensemble parameters were determined 
empirically by searching for an optimum reproduction of the measured fragment $Z$ yields and
correlations. The sensitivity
of the liquid-drop parameters of the fragment description in the hot freeze-out environment was 
studied and, as a main result, a significant reduction of the symmetry-term coefficient $\gamma$
was found necessary to reproduce the mean neutron-to-proton ratios $\langle N \rangle /Z$ and the 
isoscaling parameters of $Z \le 10$ fragments. With $\gamma = 14$~MeV a globally good reproduction 
was obtained while even lower values are suggested by some observations. Studied as a function of 
the impact parameter, $\gamma$ is found to decrease rapidly as the multiplicity of fragments and 
light particles from the disintegration of the produced spectator systems increases.
The reduction is larger than that expected from the decreasing fragment size and the 
correspondingly larger surface effects as modeled with standard mass formulas. 
It is interpreted as resulting from the overall 
reduced density of the system at breakup and the modified fragment properties in the hot environment. 
Deformations and reduced fragment densities at chemical freeze-out
contribute to producing the observed yields of neutron-rich fragments which, according to the 
present investigation with the SMM, are particularly sensitive to the widths of the isotope 
distributions and thus to the symmetry-energy term at freeze-out.

The reproduction of the experimental data with the SMM ensemble calculations is found to be very
satisfactory. A global normalization on the level of the $Z_{\rm bound}$-sorted event distributions 
was found sufficient for reaching an overall quantitative description of the $Z$ and $A$ resolved 
fragment yields. 
The calculations have, therefore, also been used to address open questions regarding the isotopic
compositions and temperatures of the system at the freeze-out stage. The $N/Z$ independence of the 
breakup temperatures in the fragmentation regime, obtained from double-isotope yields of light 
fragments, is quantitatively reproduced by the microscopic SMM temperature. The good agreement 
suggests that the opening phase space is sufficient to initiate the transition from predominantly 
residue formation to multi-fragment breakups as the transferred energy increases. Changes of the 
isotopic evolution of the spectator system between its formation during the initial nucleon-nucleon 
cascades and its subsequent breakup have been ruled out as alternative explanations for the 
isotopic fragment compositions. A reduced difference of the system $N/Z$ at breakup will result in 
smaller isoscaling parameters but, at the same time, require even larger reductions of the 
symmetry-term coefficient in order to reproduce the mean $\langle N \rangle /Z$ of intermediate-mass
fragments. 

It is finally emphasized that the simultaneous analysis of several observables obtained from a 
nearly complete coverage of the spectator kinematic regime has been important for reaching the 
presented conclusions.
The deduced modifications of hot fragments in the low-density freeze-out environment and in the 
presence of other fragments and nuclei can be expected to influence the modeling of  
astrophysical processes.

\begin{acknowledgments}

The authors gratefully acknowledge the contributions of the GSI accelerator division in
providing high-intensity and high-quality $^{124}$Sn and $^{142}$Nd beams and technical support.
R. O. thanks TUBITAK and DFG for financial support.
A.S.B. is grateful to FIAS Frankfurt, the GSI Helmholtzzentrum Darmstadt, and the IKP of Mainz
University for support and hospitality.
C.Sf. acknowledges support by the Alexander-von-Humboldt Foundation.
This work has been supported by the European Community under Contract No. HPRI-CT-1999-00001 and 
by the Polish Ministry of Science and Higher Education under Contracts No. 1 P03B 105
28 (2005 - 2006) and No. N202 160 32/4308 (2007-2009), and it has been partly supported by Grants 
NS-7235.2010.2 and RFBR 09-02-91331 (Russia).

\end{acknowledgments}

\end{document}